\documentclass[prl,%
 reprint,
 amsmath,amssymb,
 aps,
]{revtex4-2}

\usepackage{graphicx}
\usepackage{dcolumn}
\usepackage{amsmath,amssymb}
\usepackage{bm}
\usepackage{braket}
\usepackage{amsmath, amssymb, physics, graphicx}

\begin{document}
\title{Theory of Supercritical Coupling And Generalized Bound States in the Continuum}
\author{Sergio Balestrieri$^\dagger$}
\affiliation{Institute of Applied Sciences and Intelligent Systems, National Research Council, Via Pietro Castellino 111, Naples, 80131, Italy}
\author{Bruno Miranda$^\dagger$}
\affiliation{Institute of Applied Sciences and Intelligent Systems, National Research Council, Via Pietro Castellino 111, Naples, 80131, Italy}
\author{Silvia Romano}
\affiliation{Institute of Applied Sciences and Intelligent Systems, National Research Council, Via Pietro Castellino 111, Naples, 80131, Italy}
\author{Gianluigi Zito}
\email{Corresponding author email: gianluigi.zito@cnr.it\\
$^\dagger$ These authors contributed equally.}

\affiliation{Institute of Applied Sciences and Intelligent Systems, National Research Council, Via Pietro Castellino 111, Naples, 80131, Italy}
           


\date{\today}

\begin{abstract}
\noindent 
Bound states in the continuum (BICs) arise from destructive interference suppressing radiation despite spectral overlap with the continuum. Here we show that Friedrich--Wintgen interference naturally emerges from a bright--dark supermode decomposition of resonances coupled through a shared radiation channel. In this basis, any finite leakage of a quasi-BIC induces a causality-driven reactive coupling enabling non-Hermitian pumping of the dark sector. We derive the optimal condition for this process and show that it corresponds to the supercritical coupling regime previously identified in \textit{Nature} 626, 765 (2024), while naturally recovering universal quasi-BIC asymmetry scaling.
Extending the theory to Dirac-like dispersions in photonic crystal slabs, we identify an open-Dirac singularity where the Dirac gap matches the supercritical regime. A four-wave Hamiltonian quantitatively reproduces rigorous coupled-wave analysis, revealing the breakdown of conventional critical coupling. Near this regime, absorptive cross-coupling induces coherent absorption interference and enables suppression of effective dissipative losses beyond conventional material limits. These results motivate the concept of a generalized bound state in the continuum (gBIC) as a limiting non-Hermitian state where  radiative and effective gain compensate, producing a true divergence of the total quality factor. Overall, this work establishes a unified framework connecting BIC interference, Dirac topology, and non-Hermitian physics for ultra-high-Q enhancement and loss engineering in open photonic systems.
\end{abstract}

 
\maketitle
\section{Introduction}

\noindent Bound states in the continuum (BICs) are non-radiative eigenstates embedded in the radiation spectrum \cite{Stillinger1975,Friedrich1985,Zhen2014,Kodigala2017,Koshelev2018,Zito2019, zhang2025universal}. In photonic crystal slabs, they are commonly classified as parametric, symmetry-protected, or Friedrich--Wintgen (FW) BICs. The latter arise from destructive interference between leaky modes coupled to a common radiation channel. Recent work has shown that these categories can be unified within a generalized FW picture \cite{Hu2022}. Symmetry-protected BICs occur at high-symmetry points, whereas parametric BICs result from interference with slowly varying backgrounds. Dirac degeneracies, i.e., linearly dispersing mode crossings protected by symmetry, form a particularly important subclass, but their connection to FW interference remains only partially developed despite growing experimental evidence \cite{Romano2020,Contractor2022,DeTommasi2023,Zito2024}.\\
\indent We recently introduced the concept of supercritical coupling, where near-field interaction between bright and dark modes enables efficient excitation of nominally non-radiative states beyond the limits of conventional critical coupling \cite{Schiattarella2024}. In systems coupled to a shared radiation channel, energy conservation enforces a coherent interaction between radiative and subradiant components. This allows dark-mode population even when the direct radiative leakage of the dark mode is vanishingly small, and naturally connects BIC physics to non-Hermitian interference phenomena governed by non-orthogonal eigenvectors \cite{Suh2004,Yang2014}.\\
\indent Here, we develop a first-principles theory of this regime. We show that the FW condition appears as a singular point in the bright--dark supermode basis, where the dark state is exactly decoupled from radiation. Any realistic departure from this ideal condition---through asymmetry, finite in-plane momentum, detuning, or leakage---generates a finite reactive coupling \(\kappa'\), constrained by Kramers--Kronig causality. This coupling mediates energy transfer from the bright sector into the dark one and leads to the optimal condition
\[
\kappa_{\mathrm{sc}}=\sqrt{\Gamma_b\Gamma_d},
\]
which defines supercritical coupling as the non-Hermitian analogue of impedance matching for high-\(Q\) quasi-BICs. Within the same framework, the standard quasi-BIC scaling \(\gamma_{\mathrm{qBIC}}\propto\alpha^2\) follows directly from the mixing angle between bright and dark supermodes.\\
\indent We then extend the theory to Dirac-like dispersions in photonic crystal slabs and metasurfaces. We identify an \emph{open-Dirac singularity}, defined by the condition that the Dirac gap becomes commensurate with the supercritical coupling scale. At this point, the quasi-BIC is optimally pumped at a finite in-plane momentum and the system approaches a ring of exceptional points (EPs) \cite{zhen2015spaw}. A four-wave Hamiltonian captures the experimentally relevant TE-like response and quantitatively reproduces RCWA simulations, including the emergence of strong modal non-orthogonality.\\
\indent The resulting regime departs sharply from conventional critical coupling. RCWA shows that the angle of maximal enhancement is no longer governed by \(Q_{\mathrm{rad}}=Q_{\mathrm{abs}}\), but by the supercritical condition. Moreover, the effective absorption of the quasi-BIC is not fixed by the material extinction coefficient alone. Absorptive cross-coupling produces a coherent absorption interference driving the spatial profile reconfiguration of the quasi-BIC. The dark radiative mode (quasi-BIC) becomes the analog \textit{bright} version with regard to the absorption loss channel, while the bright radiative mode becomes \textit{dark} to absorption as confining in void regions. Concurrently, the global quasi-BIC field delocalization, though,  reduces the energy fraction stored within the lossy slab, inducing a diverging scaling of  \(Q_{\mathrm{abs}}^{\mathrm{eff}}\). These findings motivate the concept of a \emph{generalized bound state in the continuum} (gBIC), a limiting non-Hermitian state in which radiative and nonradiative balance leads to a zero total decay rate state in parity-time $(\mathcal{PT})$ symmetric photonic resonators, i.e.  with diverging  \textit{total} quality factor in the continuum.\\
\indent Our results unify BIC interference, Dirac topology, and non-Hermitian physics, providing a general framework for engineering radiation, loss, and field enhancement in open photonic systems, with direct implications for Dirac lasing and EP-based devices \cite{chua2014,gao2020,Contractor2022,ma2023,masharin2023}.\\
\indent \indent The paper is organized as follows. We first formulate the shared-radiation-channel problem and introduce the bright--dark supermode basis, showing how the Friedrich--Wintgen condition emerges as the singular BIC limit (Sec. II). We then derive the quasi-BIC asymmetry scaling (Sec. III) and use Kramers--Kronig causality to show why finite leakage necessarily generates reactive coupling (Sec. IV). Next, we establish the supercritical condition for optimal dark-mode pumping (Sec. V) and extend the theory to Dirac dispersions, identifying the open-Dirac singularity (Sec. VI) and its connection to exceptional-point rings (Sec. VII). Finally, we validate the model by RCWA, analyze the breakdown of conventional critical coupling (Sec. VIII), and show how absorption interference (Sec. IX) and field delocalization (Sec. X) motivate the generalized-BIC concept (Sec. XI).

\section{Shared radiation channel}
 \noindent The response of open photonic resonators can be rigorously formulated within a non-Hermitian Hamiltonian framework (see Supporting Information, Sec.~I for a detailed derivation). We begin by analyzing a minimal system of two coupled resonances interacting through a single radiation channel, representative of one-dimensional gratings. This framework is then extended to two-dimensional photonic crystal slabs and metasurfaces, where three- and four-wave interactions naturally arise. Assuming a harmonic time dependence $e^{i\omega t}$, we investigate both the dynamical and steady-state behavior of the system.
The two modes are represented by the amplitude vector $\mathbf{a}(t) = [x_1(t), x_2(t)]^T$, and are coupled to an external radiation channel $s^+ \rightarrow s^-$, and to each other via near-field coupling $\kappa$. The modal amplitudes must be time-dependent due to their inherent decay into the environment. The temporal coupled-mode equations governing their evolution in a lossless system under time-reversal symmetry are given by \cite{Suh2004}:
\begin{align}
\frac{d}{dt}{\mathbf{a}} &= \left(i\Omega-\Gamma_{\rm rad}\right)\mathbf{a}+\mathbf{K}s_+,\\
s_- &= C\,s_+ + \mathbf{D}^T\mathbf{a}.\label{eq:mainTCMT}
\end{align}
where $\Omega$ include modal frequencies $\omega$ and reactive coupling $\kappa$,
\[
\Omega = \begin{bmatrix} \omega_1 & \kappa \\ \kappa & \omega_2 \end{bmatrix}.
\]
Here $\mathbf{K}$ is the input-coupling vector, $\mathbf{D}$ the output-coupling vector, and $C$ the direct scattering coefficient, with $s^+$ denoting the incoming wave and $s^-$ the outgoing wave.
Normalizing the stored energy as $\mathcal{E}=\mathbf{a}^\dagger\mathbf{a}$ and the input/output powers as $P_\pm=|s_\pm|^2$, energy conservation requires
\begin{equation}
\frac{d}{dt}\left(\mathbf{a}^\dagger\mathbf{a}\right)=|s_+|^2-|s_-|^2.
\label{eq:power_bd_main}
\end{equation}
Substitution yields the standard constraints:
\begin{align}
\Gamma_{\rm rad} &= \frac{1}{2}\mathbf{D}\mathbf{D}^\dagger, \label{eq:Gamma_bd_constraint_main}\\
\mathbf{K} &= \mathbf{D},\\
C &= -1.
\end{align}
In real systems, nonradiative losses can be included as $\Gamma_{\text{tot}} = \Gamma_{\text{rad}} + \Gamma_{\text{abs}}$, representing the total decay matrix, including modes-1,2 radiation $\gamma_{1,2}$ and dissipative far field radiation coupling $\gamma_{12}$:
\[
\Gamma_{\text{rad}}= \begin{bmatrix} \gamma_1 & \gamma_{12} \\ \gamma^{\star}_{12} & \gamma_2 \end{bmatrix}.
\]
The structure of the absorption operator $\Gamma_{\text{abs}}$ will be analyzed in detail in a later section, where absorption-induced interference and modal non-orthogonality are explicitly considered. Here, it is assumed to be diagonal. This approximation corresponds to neglecting absorptive cross-coupling terms, which can always be eliminated to first order by a unitary rotation of the modal basis when absorption is weak and spatially uniform, leaving only diagonal (mode-independent) loss.

The radiation amplitudes of the two modes can be expressed as:
\[
\mathbf{q} = \begin{bmatrix} \sqrt{\gamma_1} e^{i\phi_1} \\ \sqrt{\gamma_2} e^{i\phi_2} \end{bmatrix},
\]
which allows us to construct the defective radiation matrix as:
\[
\Gamma_{\text{rad}} = \mathbf{q} \mathbf{q}^\dagger,
\]
implying it is a rank-1 Hermitian positive-semidefinite matrix, appropriate for a single shared radiation channel.  The phase difference of radiated waves $\theta = \phi_1 -\phi_2$ depends on mode symmetry \cite{Suh2004}.\\
\indent It must be noted that energy conservation is guaranteed only if $\Gamma_{\text{rad}}$ is positive semi-definite, and particularly if the off-diagonal elements are non-zero. Otherwise, power conservation in the shared radiation channel is violated, as we will see explicitly in Section V.\\
\indent We seek a unitary matrix $U = [\mathbf{v}_b, \mathbf{v}_d]$  such that:
\[U^\dagger \Gamma_{\text{rad}} U = \text{diag}(\gamma, 0).\]
The basis vectors defining bright and dark supermodes are explicitly:
\begin{align*}
\mathbf{v}_b &= \frac{1}{\sqrt{\gamma_1 + \gamma_2}} \begin{bmatrix} \sqrt{\gamma_1} e^{i\phi_1} \\ \sqrt{\gamma_2} e^{i\phi_2} \end{bmatrix}, \\
\mathbf{v}_d &= \frac{1}{\sqrt{\gamma_1 + \gamma_2}} \begin{bmatrix} \sqrt{\gamma_2} e^{-i\phi_2} \\ -\sqrt{\gamma_1} e^{-i\phi_1} \end{bmatrix}.
\end{align*}
Thereby, we have:
\[U = \frac{1}{\sqrt{\gamma_1 + \gamma_2}} \begin{bmatrix} \sqrt{\gamma_1} e^{i\phi_1} & \sqrt{\gamma_2} e^{-i\phi_2} \\ \sqrt{\gamma_2} e^{i\phi_2} & -\sqrt{\gamma_1} e^{-i\phi_1} \end{bmatrix},\] 
which rotates the system into the bright/dark basis, diagonalizing \( \Gamma_{\text{rad}} \) into:
\[
U^\dagger \Gamma_{\text{rad}} U = \begin{bmatrix} \gamma_b & 0 \\ 0 & \gamma_d \end{bmatrix},
\quad \text{with} \quad \gamma_b = \gamma_1 + \gamma_2, \quad \gamma_d = 0.
\]
The orthogonal dark mode vector satisfies $\mathbf{q}^\dagger \cdot \mathbf{v}_d = 0$. Any radiating mode must therefore be a linear combination of these basis vectors.
\\
\indent Next, rotating the Hermitian part $\Omega$ into the bright/dark basis yields:
\[
\Omega' = U^\dagger \Omega U = \begin{bmatrix} \omega_b & \kappa' \\ \kappa'^* & \omega_d \end{bmatrix},
\]
with explicit terms:
\begin{align}
\omega_b &=  
\mathbf{v_b^\dagger}
 \Omega\mathbf{v_b} = \frac{\gamma_1 \omega_1 + \gamma_2 \omega_2 + 2\kappa \sqrt{\gamma_1\gamma_2} \cos\theta}{\gamma_1 + \gamma_2}, \\
\omega_d &=  
\mathbf{v_d^\dagger}
 \Omega\mathbf{v_d} = \frac{\gamma_2 \omega_1 + \gamma_1 \omega_2 - 2\kappa\sqrt{\gamma_1\gamma_2}  \cos\theta}{\gamma_1 + \gamma_2}, \\
\kappa' &=  
\mathbf{v_b^\dagger}
 \Omega\mathbf{v_d} = \frac{(\omega_1 - \omega_2)\sqrt{\gamma_1\gamma_2} e^{i\theta} - \kappa(\gamma_1 - \gamma_2)}{\gamma_1 + \gamma_2}.
\end{align}
These expressions describe the emergence of coupled-resonator induced transparency  at the frequency $\omega_d$. See as instance the specific case $\kappa = 0$ and $\theta = 0$ in ref. \cite{hsu2014theo}: when a weakly radiating (dark) mode couples to a strongly radiating (bright) one, a sharp dispersion arises, slowing down light and enabling field buildup — a form of open-cavity enhancement \cite{Yang2014}. \\
\indent Requiring $\kappa' = 0$ to diagonalize $\Omega'$ corresponds to:
\begin{equation}
 \boxed{ \kappa'=0}\;\; \Leftrightarrow \;\;
 \boxed{\kappa (\gamma_1 - \gamma_2) = (\omega_1 - \omega_2)\sqrt{\gamma_1 \gamma_2}e^{i\theta}},\label{eq:FW}  \end{equation}
which is precisely the Friedrich--Wintgen condition for BIC formation. The dark mode becomes a true embedded eigenstate with \emph{zero} radiative decay, effectively rendering the eigenvalue real with no far-field decay.  The Friedrich–Wintgen condition represents a singular limit of supermode interference. In the bright/dark basis $\mathbf{a} = [x_b, x_d]^T$ with driving $\mathbf{K}s_+ = [f_b(t), 0]^T$, Eq.~(1) reads:
\begin{eqnarray}
\dot{x}_b &=& i\omega_b x_b - \gamma_b x_b + i\kappa' x_d + f_b(t),\\
\dot{x}_d &=& i\omega_d x_d + i\kappa'^* x_b.
\end{eqnarray}
The only coupling between $x_b$ and $x_d$ is the off-diagonal term $\kappa'$. If $\kappa'=0$, the dark mode decouples, $\dot{x}_d = i\omega_d x_d$, yielding no energy exchange or amplitude growth (as evident in the rotating frame $\omega_d = 0$). This singular limit is consistent with energy conservation, as only the bright mode radiates. 
In realistic systems, however, a quasi-dark mode always possesses finite decay, which necessarily implies $\kappa' \neq 0$. Although the non-Hermitian Hamiltonian can be diagonalized, its eigenvectors become non-orthogonal in a single radiation channel \cite{Suh2004}. When expressed back in the orthogonal bright/dark basis, this non-orthogonality manifests as an effective reactive coupling $\kappa'$, encoding the underlying modal interference. This coupling vanishes only in the ideal Friedrich–Wintgen limit.\\
\indent In the following, we derive $\kappa'$ from radiation leakage, establish the supercritical condition, and determine its constraints from causality.

\section{Asymmetry scaling as radiation leakage}

\noindent Let $[x_1, x_2]^T$ denote two coupled resonators. Any finite radiation can be written as a rank-1 matrix:
\begin{eqnarray}
\Gamma_{\text{rad}} = \gamma_b \,\mathbf{u}\mathbf{u}^\dagger 
= \gamma_b
\begin{bmatrix}
\cos^2\Phi & \cos\Phi\sin\Phi \\
\cos\Phi\sin\Phi & \sin^2\Phi
\end{bmatrix},
\end{eqnarray}
where $\Phi$ defines the mixing with the radiative channel.
For $\Phi=0$, $\Gamma_{\text{rad}}$ commutes with the parity operator 
$P=\begin{bmatrix}1 & 0 \\ 0 & -1\end{bmatrix}$, and bright/dark modes are symmetry-protected. For $\Phi \neq 0$, the off-diagonal terms $\sim \cos\Phi\sin\Phi$ break this commutation, $[P,\Gamma_{\text{rad}}]\neq 0$, signaling inversion-symmetry breaking.
The corresponding supermodes are
\[
x_b = \cos\Phi\, x_1 + \sin\Phi\, x_2, \quad
x_d = -\sin\Phi\, x_1 + \cos\Phi\, x_2,
\]
for which $\Gamma'_{\text{rad}} = \mathrm{diag}(\gamma_b,0)$.  
When $\Phi \neq 0$, the dark mode acquires finite radiation, becoming a quasi-BIC with leakage:
\[
\gamma_{\text{qBIC}} = \gamma_b \sin^2\Phi \approx \gamma_b \Phi^2 \quad (\Phi \ll 1).
\]
This directly recovers the universal scaling $\gamma_{\text{qBIC}} \propto \alpha^2$ \cite{Koshelev2018}, with $\alpha \propto \Phi$. Hence, quasi-BIC radiation arises from a rotation of the symmetry-protected dark state, with leakage set by symmetry breaking.
In the original basis $\mathbf{a} = [x_1,x_2]^T$, a finite $\Phi$ implies $\gamma_1 \neq \gamma_2$, which can only occur via geometric asymmetry or finite in-plane momentum; at the $\mathbf{\Gamma}$ point with inversion symmetry, one must have $\gamma_1 = \gamma_2$.

\section{Kramers--Kronig causality}

\noindent To establish when a finite reactive coupling $\kappa'$ must arise, we recall the constraints on the radiation operator encoded in Eq.~(\ref{eq:Gamma_bd_constraint_main}).
In the bright/dark basis, the radiative decay matrix is a positive-semidefinite rank-1 matrix.
For two modes coupled to the same radiation channel,
\begin{equation}
\mathbf{D}=
\begin{pmatrix}
d_b\\ d_d
\end{pmatrix},
\qquad
\Gamma_{\rm rad}
=
\frac{1}{2}
\begin{pmatrix}
|d_b|^2 & d_b d_d^\ast\\
d_d d_b^\ast & |d_d|^2
\end{pmatrix}.
\label{eq:D_Gamma_main}
\end{equation}
Hence
\begin{equation}
\gamma_b^{\rm rad}=\frac{1}{2}|d_b|^2,\qquad
\gamma_d^{\rm rad}=\frac{1}{2}|d_d|^2,\qquad
\Gamma_{bd}=\frac{1}{2}d_b d_d^\ast.
\end{equation}
We identify the diagonal elements as the radiative decay rates of the bright and dark modes,
\begin{equation}
\gamma_b^{\rm rad} = \frac{1}{2}|d_b|^2,\qquad
\gamma_d^{\rm rad} = \frac{1}{2}|d_d|^2,
\end{equation}
and the off-diagonal element as
\begin{equation}
\Gamma_{bd} = \frac{1}{2}d_b d_d^\ast.
\end{equation}
In the \emph{ideal} limit, the \textit{purely} dark mode does not radiate at all:
\begin{equation}
\gamma_d^{\rm rad}=0 \quad\Longleftrightarrow\quad d_d=0.
\end{equation}
Then
\begin{equation}
\Gamma_{\rm rad}^{(\rm FW)} =
\frac{1}{2}
\begin{pmatrix}
|d_b|^2 & 0\\[3pt]
0 & 0
\end{pmatrix},
\end{equation}
which is diagonal in the bright/dark basis and has rank 1. 

Now suppose that due to geometric imperfections, finite in-plane momentum, or symmetry breaking, the dark mode acquires \emph{finite} radiative leakage into the same channel, i.e.
\begin{equation}
\gamma_d^{\rm rad} > 0 \quad\Longleftrightarrow\quad d_d \neq 0,\qquad d_b\neq 0
\end{equation}
(the bright mode remains radiative). Then, from Eq.~\eqref{eq:D_Gamma_main},
\begin{equation}
\Gamma_{bd} = \frac{1}{2}d_b d_d^\ast \neq 0.
\end{equation}
Hence,  any \textit{finite radiative leakage} of the \textit{quasi} dark physical mode into the shared channel immediately induces a nonzero \textit{radiative cross-coupling} $\Gamma_{bd}$.

The matrix $\Gamma_{\rm rad}(\omega)$ encodes the \emph{on-shell} radiative decay at frequency $\omega$, arising from coupling to a continuum of radiation states. This coupling is described by a retarded self-energy
\begin{equation}
\Sigma(\omega) = \Delta(\omega) + \frac{i}{2}\Gamma_{\rm rad}(\omega),
\end{equation}
where $\Delta(\omega)$ is the reactive Lamb-shift part. Causality implies that $\Sigma(\omega)$ is analytic in the upper half-plane, so $\Delta(\omega)$ and $\Gamma_{\rm rad}(\omega)$ are related by Kramers–Kronig relations. In particular, the off-diagonal elements obey
\begin{equation}
\Delta_{bd}(\omega) =
\frac{1}{\pi}\,\mathcal{P}\!\!\int_{-\infty}^{\infty}
\frac{\Gamma_{bd}(\omega')}{\omega'-\omega}\,d\omega'.\label{eq:KK}
\end{equation}
Thus, if $\Gamma_{bd}(\omega)$ is nonzero over any finite bandwidth, its Hilbert transform $\Delta_{bd}(\omega)$ cannot vanish identically. Except for finely tuned cancellations, a nonzero radiative cross-coupling $\Gamma_{bd}(\omega)$ \emph{forces} a nonzero reactive cross-coupling $\Delta_{bd}(\omega)$.

The effective non-Hermitian Hamiltonian in is
\begin{equation}
H_{\rm eff}(\omega) = \Omega + \Delta(\omega) + i\,\Gamma_{\rm rad}(\omega),
\end{equation}
so off-diagonal entries are
\begin{equation}
H_{{\rm eff},bd}(\omega) = \Omega_{bd} + \Delta_{bd}(\omega) + i \Gamma_{bd}(\omega).
\end{equation}
Even if the bare Hermitian coupling $\Omega_{bd}$ were zero at the FW point, the self-energy correction $\Delta_{bd}(\omega)\neq0$ induced by the shared radiation channel, necessary in any physical  rotated version of ideal bright and dark modes,
enables the near-field mixing that underlies non-Hermitian pumping in realistic quasi-BIC systems.

\textbf{Material Absorption.} All possible perturbations place every physical quasi-BIC in a neighbourhood of the FW point, also material absorption.\\
\indent Let the absorption matrix for waves $[x_1,x_2]^{T}$ be diagonal but mode-asymmetric:
\begin{equation}
\Gamma_{\mathrm{abs}}(\omega)=
\begin{bmatrix}
\delta_1(\omega) & 0\\[3pt]
0 & \delta_2(\omega)
\end{bmatrix},
\qquad 
\Delta\delta(\omega)=\delta_2(\omega)-\delta_1(\omega).
\end{equation}
Transforming to the bright/dark basis with the unitary rotation $U$ yields
\begin{equation}
\Gamma_{\mathrm{abs}}^{(bd)} = U^\dagger\Gamma_{\mathrm{abs}}U 
=
\frac{1}{2}
\begin{pmatrix}
\delta_1+\delta_2 & \Delta\delta \\[3pt]
\Delta\delta & \delta_1+\delta_2
\end{pmatrix}.
\end{equation}
This asymmetry, $\Delta\delta\neq0$, produces a \emph{dissipative} off-diagonal term $\Gamma_{bd}=\Delta\delta/2$. Because absorption is a causal response function, the asymmetry \emph{must} generate an associated \emph{reactive} coupling through Eq.\eqref{eq:KK}.
Physically, the Lamb shift $\Delta_{bd}(\omega)$ represents a coherent frequency renormalization arising from the same microscopic processes that generate absorption. \\
\indent The absorption profile in photonic materials is rarely flat; it arises from electronic transitions, defect-assisted continua, excitonic tails, or phonon bands. Each exhibits a Lorentzian form near resonance, which we model as
\begin{equation}
\delta_{1,2}(\omega)=\delta_{1,2}^{(0)}+
\frac{A_{1,2}}{1+\big[(\omega-\omega_a)/\Lambda\big]^2},
\end{equation}
where $\omega_a$ is the microscopic resonance frequency and $\Lambda$ its linewidth. The resulting asymmetry,
\begin{equation}
\Delta\delta(\omega)=\Delta\delta^{(0)}+
\frac{\Delta A}{1+\big[(\omega-\omega_a)/\Lambda\big]^2},
\end{equation}
feeds Kramers–Kronig and produces the reactive correction
\begin{equation}
\Delta_{bd}(\omega)=
\frac{\Delta A}{2\Lambda}\,
\frac{\omega-\omega_a}{1+\big[(\omega-\omega_a)/\Lambda\big]^2}.
\end{equation}
The Lamb shift therefore inherits the same linewidth $\Lambda$ as the absorptive resonance. It changes sign across $\omega=\omega_a$, grows with $\Delta A$, and is maximal at $|\omega-\omega_a|=\Lambda$.\\
\indent The effective off-diagonal coupling is the sum of radiative-, geometric- and absorption-induced terms:
\begin{equation}
\kappa'_{\mathrm{eff}}(\omega)=
\kappa_{\mathrm{geo}}
+\kappa_{\mathrm{rad}}
+\Delta_{bd}(\omega).
\end{equation}
Here $\kappa_{\mathrm{geo}}$ arises from structural asymmetry, and $\kappa_{\mathrm{rad}}$ appears only if the (quasi) dark mode acquires radiative leakage (or finite $k_{\parallel} - k_{\rm{BIC}}\neq 0$). When the system is close to the FW-BIC limit, $\kappa_{\mathrm{rad}}$ is extremely small and the dominant correction comes from either $\kappa_{\mathrm{geo}}$ or $\Delta_{bd}(\omega)$. 
Following the definition of supercritical coupling described in next Sec. V, numerical estimates showing that realistic absorption asymmetries and modest geometric perturbations can generate such feasible regime are given in  Supporting Information, Sec.~S2. 
\begin{figure*}[t!]
    \centering
    \includegraphics[width=1\linewidth]{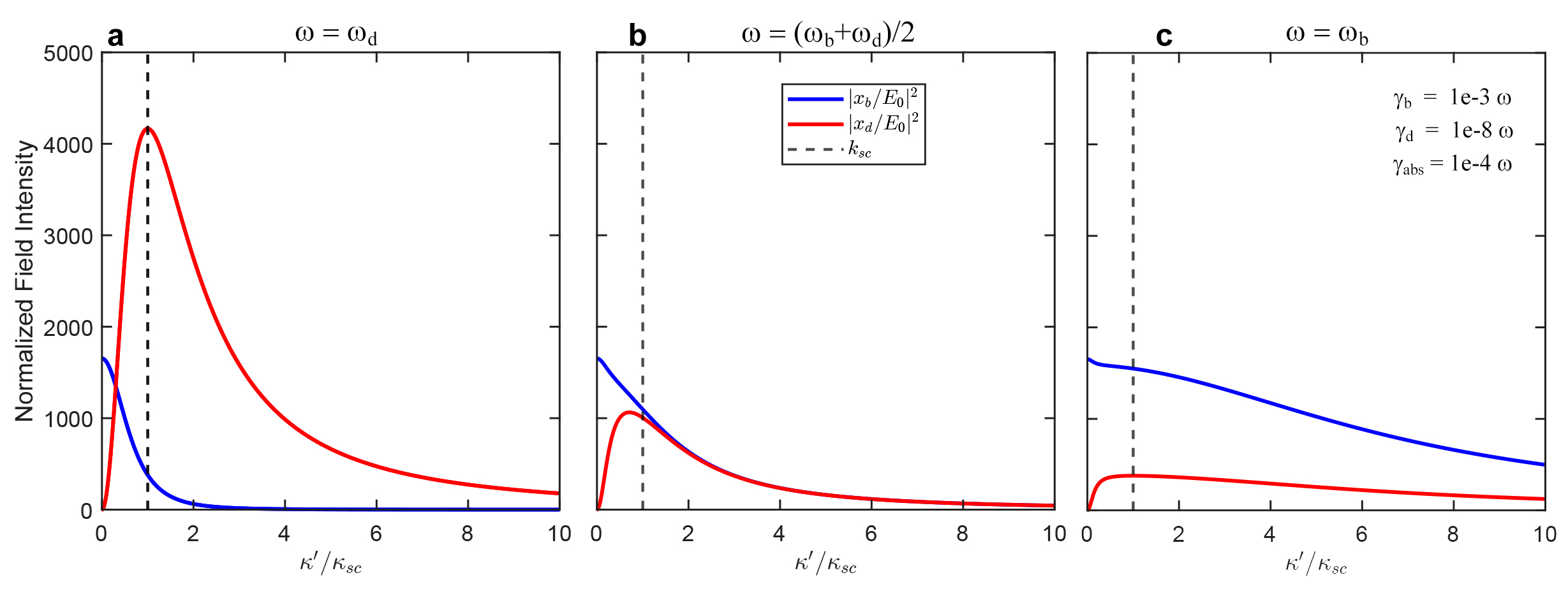}
    \caption{Dynamic evolution of dark and bright mode populations under continuous wave drive at frequency $\nu = \omega/2\pi= 299$ THz ($\lambda = 1 \mu$m), for different values of $\kappa'$ around $\kappa_{sc}$, and for fixed parameters of radiative and non radiative losses as displayed.}
    \label{fig:1}
\end{figure*}

\begin{figure*}
    \centering
    \includegraphics[width=1\linewidth]{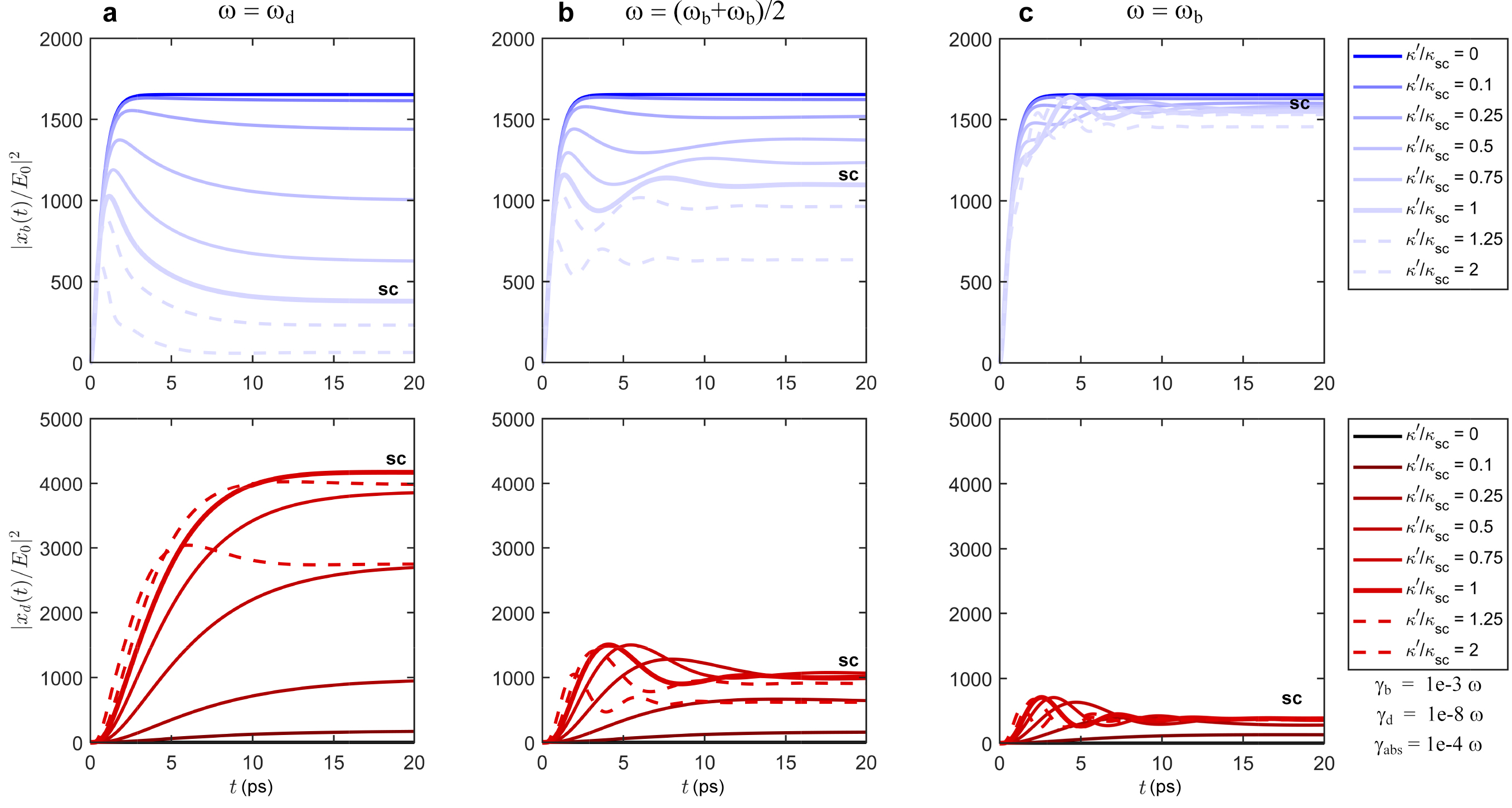}
    \caption{Dynamic evolution of dark and bright mode populations under continuous wave drive at frequency $\omega$, for different values of $\kappa'$ around $\kappa_{sc}$, and for fixed parameters of radiative and non radiative losses as displayed.}
    \label{fig:3}
\end{figure*}

\begin{figure}
    \centering
    \includegraphics[width=1\linewidth]{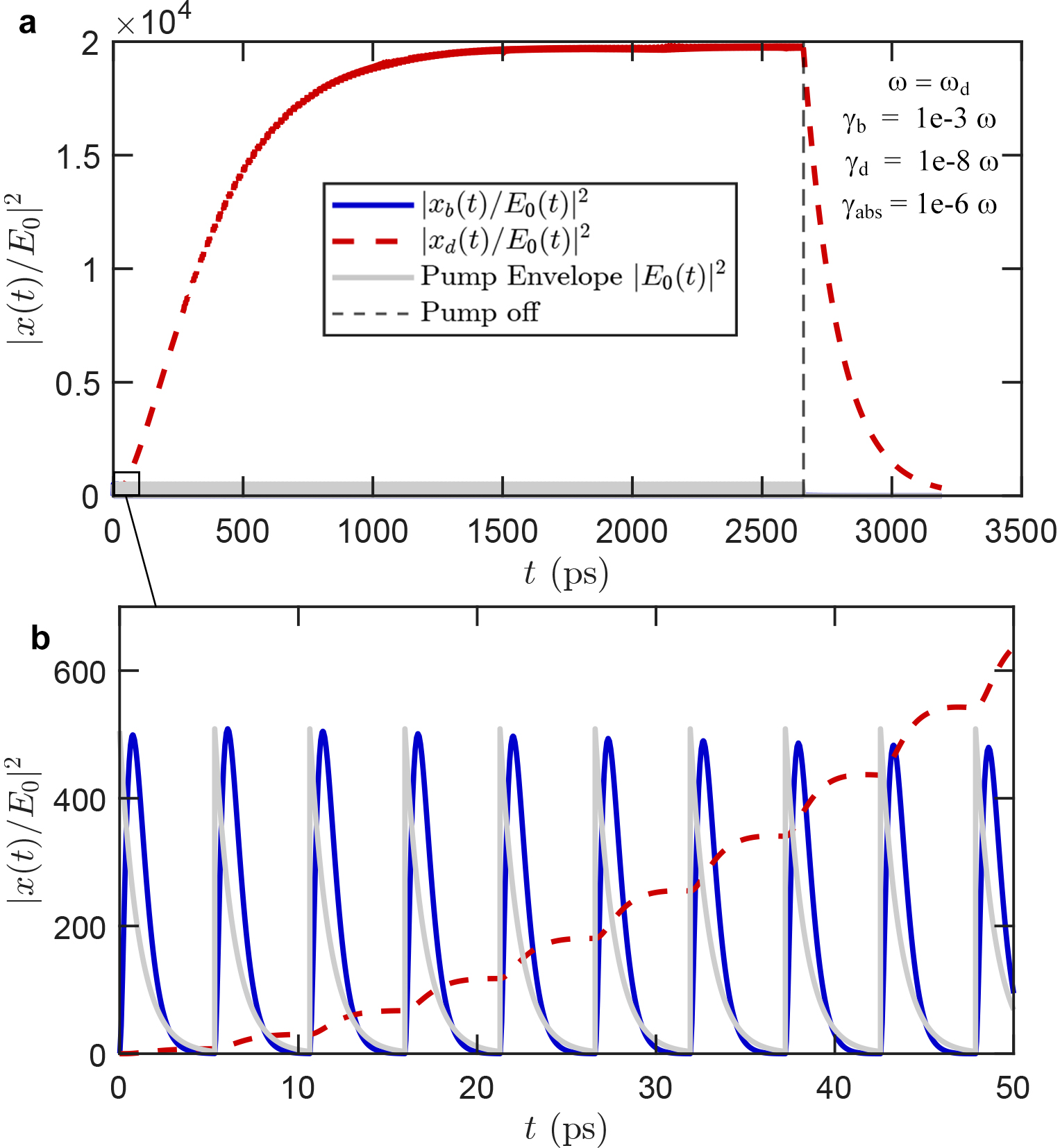}
    \caption{Dynamics of dark and bright mode populations under pulsed wave excitation at frequency $\nu = \omega/2\pi= 299$ THz ($\lambda = 1\, \mu$m), at $\kappa'=\kappa_{sc}$, for fixed decay parameters as displayed. The population of the dark mode saturates at a level imposed by $\gamma_{\text{abs}}$, which is reached with a characteristic time $\sim1/\gamma_{\text{abs}}$. The pulse maximum is normalized to the bright mode peak at $t = 0$, for visualization. The bright mode maximum decays with time stabilizing at its steady state value after a sufficient number of pulses. As the train of pulses (500) stops, the dark mode decays obeying $\gamma_{\text{abs}} = 10^{-6}\omega_d$ (nanosecond time scale) since $\gamma_d \ll\gamma_{\text{abs}} $.}
    \label{fig:4}
\end{figure}


\section{Supercritical coupling condition as optimal reactive coupling}
\noindent We now consider small deviations from the ideal BIC condition---arising from momentum shift, asymmetry, detuning, or weak radiation leakage induced by non-ideal perturbations. In the bright/dark basis, this introduces a finite $\gamma_d$ and, by causality, a non-zero reactive coupling $\kappa'$, as discussed in the previous section. The interaction is therefore governed by the effective Hamiltonian
\begin{equation}
H_{\text{eff}}=
\begin{bmatrix}
\omega_b+i\gamma_b & \kappa'\\
\kappa'^* & \omega_d+i\gamma_d
\end{bmatrix}.
\end{equation}
The stationary coupled-mode equations, including material absorption $\gamma_{\text{abs}}$, read
\begin{equation}
\left(
\begin{bmatrix}
\omega_0+i\Gamma_b & \kappa'\\
\kappa' & \omega_0+\Delta+i\Gamma_d
\end{bmatrix}
-\omega I
\right)
\begin{bmatrix}
x_b\\
x_d
\end{bmatrix}
=
i
\begin{bmatrix}
gE_0\\
0
\end{bmatrix},
\end{equation}
where $f_b=gE_0$, with $g=\sqrt{2\gamma_b}$ for time-reversal-symmetric coupling, and $E_0$ is the amplitude of the external driving field $E(t)$, whose temporal envelope will be specified later. For simplicity, direct driving of the dark mode is neglected. The total losses are $\Gamma_{b,d}=\gamma_{b,d}+\gamma_{\text{abs}}$, with $\gamma_d\ll\gamma_b$. Here $\omega_0$ is the resonance frequency and $\Delta$ the detuning between bright and dark modes. We first assume zero detuning, $\Delta=0$, with the pump resonant with the dark mode and $\kappa'\ll\gamma_b$.
Solving for the stationary amplitudes gives
\begin{equation*}
x_b=\frac{gE_0}{\Gamma_b+\frac{\kappa'^2}{\Gamma_d}},
\end{equation*}
and
\begin{equation*}
x_d=\frac{i\kappa'}{\Gamma_d}\cdot\frac{gE_0}{\Gamma_b+\frac{\kappa'^2}{\Gamma_d}}
=\frac{igE_0\kappa'}{\Gamma_b\Gamma_d+\kappa'^2}.
\end{equation*}
The dark-mode enhancement is therefore
\begin{equation}
\label{eq:efficiency}
\eta=\left|\frac{x_d}{E_0}\right|^2
=\frac{(g\kappa')^2}{(\Gamma_b\Gamma_d+\kappa'^2)^2}.
\end{equation}
Maximization with respect to $\kappa'$ immediately yields the supercritical coupling condition
\begin{equation}
\boxed{
\kappa_{\mathrm{sc}}
=\sqrt{\Gamma_b\Gamma_d}
=\sqrt{(\gamma_b+\gamma_{\text{abs}})(\gamma_d+\gamma_{\text{abs}})}
\simeq \sqrt{\gamma_b\gamma_{\text{abs}}}
},
\end{equation}
where the approximation holds for $\gamma_d\ll\gamma_{\text{abs}}\ll\gamma_b$. This agrees with the result in Ref.~\cite{Schiattarella2024}, although in that case direct driving of the dark mode was included, yielding a symmetric expression under $x_b\leftrightarrow x_d$.\\
\indent For finite detuning, $\Delta\neq0$, the general expression becomes
\begin{equation}
\eta(\Delta)=\left|\frac{x_d}{E_0}\right|^2
=
\frac{g^2\kappa'^2}
{\left|(\Delta+i\Gamma_d)\left(i\Gamma_b+\frac{\kappa'^2}{\Delta+i\Gamma_d}\right)\right|^2},
\end{equation}
which still yields an optimum near $\kappa_{\mathrm{sc}}\approx\sqrt{\Gamma_b\Gamma_d}$, neglecting terms of order higher than $\kappa'^4$ and provided that the drive frequency $\omega$ remains resonant with $\omega_d$.\\
\indent We now analyze the time-domain evolution of both mode populations as a function of $\kappa'$ and $\gamma_{\text{abs}}$, in order to demonstrate optimal non-Hermitian pumping under both continuous-wave and pulsed excitation. The analysis  is obtained by numerically solving Eq.~(1) using a custom finite-difference MATLAB code (MathWorks).
\indent \textbf{Figure~\ref{fig:1}} shows the maximum mode population in the steady-state regime as a function of the ratio \( \kappa'/\kappa_{sc} \), for a representative detuning of the drive frequency \( \omega \) from the resonant frequencies. Recall that increasing \( \kappa' \) leads to a greater spectral separation between eigenmodes, which accounts for the reduction in population of modes whose resonances lie farther from the pump frequency. Additionally, the effective decay rate of the dressed modes scales as \( \kappa'^2 \), implying that the maximum attainable field intensity (or population) follows this trend as \( \kappa'/\kappa_{sc} \) increases.\\
\indent Indeed, if $\Gamma_b \gg \Gamma_d, \kappa'$, then the bright mode decays quickly and follows the drive nearly instantaneously. Adiabatically eliminating $\dot{x}_b \approx 0$ from Eq. (5) and (6) in the rotating frame yields:
\begin{equation}
 \dot{x}_d = -\left( \Gamma_d + \frac{\kappa'^{2}}{\Gamma_b} \right) x_d + \frac{i \kappa'}{\Gamma_b} gE(t). 
\end{equation}
Considering This relation indicates that the dark mode decays with an enhanced effective decay rate
$\Gamma_{\text{eff,tot}} = \Gamma_d + \frac{\kappa'^{2}}{\Gamma_b}$,  
and the drive indirectly reaches the dark mode with an effective strength $\frac{\kappa'}{\Gamma_b}$.
The excess rate of decay at supercritical coupling provides a significant outcome when $\Gamma_d \to\gamma_{\text{abs}}$ (radiation loss is effectively zero and only absorption remains), indeed:
\begin{equation}
 \Gamma_{\text{eff,tot}} = \gamma_{\text{eff,rad}}+\gamma_{\text{abs}} = \Gamma_d +\frac{\kappa_{sc}^2}{\Gamma_b} \to \gamma_{\text{abs}} + \gamma_{\text{abs}}, 
\end{equation}
which immediately  imposes that the effective radiative loss of the dark mode reaching supercritical coupling is:
\begin{equation}
 \boxed{ \gamma_{\text{eff,rad}}(\kappa_{sc})=\gamma_{\text{abs}}.} 
\label{eq:eff_critical}
\end{equation}
Therefore it  is ultimately determined by the nonradiative loss, as evidenced in \textbf{Supporting Fig. 1}. Equation~\eqref{eq:eff_critical} has the \emph{form} of the usual critical-coupling condition, $\gamma_{\rm rad}=\gamma_{\rm abs}$, but here it emerges only \emph{indirectly}. The radiative loss of the quasi-dark mode is generated by hybridization of the effective state, $(1-\varepsilon)\mathbf{v}_d+\varepsilon\mathbf{v}_b$ with $\varepsilon\ll1$, and is automatically driven toward the intrinsic decay rate when $\kappa'$ is tuned to $\kappa_{\mathrm{sc}}$. In this sense, supercritical coupling realizes an effective impedance matching of the dark-like mode through non-Hermitian bright-mode admixture, rather than through direct tuning of its own radiation channel. This makes impedance matching possible even for extremely high-$Q$ quasi-BICs, where classical critical coupling would fail.\\
\indent The maximum intensity achievable in the bright and dark modes, as a result of non-Hermitian pumping at the optimal reactive coupling becomes:
\begin{eqnarray}
|x_b|^2 &=& \frac{g^2E_0^2}{4\Gamma_b^2} \simeq \frac{E_0^2}{2\gamma_b},\\
|x_d|^2 &=& \frac{g^2E_0^2}{4\Gamma_b\Gamma_d} \simeq \frac{E_0^2}{2\gamma_{\text{abs}}} \left(1-\frac{\gamma_{\text{abs}}}{\gamma_b}\right ).\label{eq:xd}
\end{eqnarray}
In case of critical coupling with the bright mode, $\gamma_{b} = \gamma_\text{abs}$, Eq.~(\ref{eq:xd}) apparently implies that energy transfer through non-Hermitian pumping to the dark mode is impeded. The exact solution instead allows dark mode population growth also in this case, as shown in \textbf{Supporting Fig. S1}.  Condition $\gamma_{b} \gg \gamma_\text{abs}$ allows the dark mode to achieve the maximum level of enhancement that a resonator can achieve, despite being an ideal trapped radiation state.\\
\indent \textbf{Figure~\ref{fig:3}} shows the mode populations under continuous-wave pumping for different parameter sets. When the pump is resonant with the dark mode, $\omega=\omega_d$, the steady-state energy stored in the bright mode drops from 1600 to 500 (\textbf{Fig.~\ref{fig:3}a}), meaning that roughly $3/4$ of the energy delivered through the bright mode is transferred to the dark mode.  In addition, is clear that even deviating from ideal $\kappa_{sc}$, the dark mode can be largely populated through reactive coupling, despite being effectively a zero radiation state in a low-loss material resonator ($\gamma_d = 10^{-8}\omega$, $\gamma_{\text{abs}} = 10^{-4}\omega$). The non-Hermitian pumping  diverts the excess of energy accessing the system through the bright mode towards the dark mode via reactive coupling, until saturation ruled by nonradiative loss is reached. When the pump is tuned on the central frequency $(\omega_b + \omega_d)/2$, the dark mode still reaches a level of field enhancement comparable with the bright mode (\textbf{Fig.~\ref{fig:3}b}), and a non-negligible  fraction of it when pump is resonant with the bright mode (\textbf{Fig.~\ref{fig:3}c}).\\
\indent The model predicts the optimal pulse decay time for accelerated dark-mode buildup, $\tau_{\text{opt}} = 1/\gamma_{\text{abs}}$,
diverging for an unphysical system with zero absorption. This picture is clearer under pulse excitation drive. \textbf{Figure~\ref{fig:4}} shows the evolution of the mode population under a train of 500 pulses, each 5 ps long. The bright mode is populated and depleted on the short timescale $\sim1/\gamma_b$, while continuously transferring part of its energy to the dark mode. The latter, with lifetime $1/\gamma_{\text{abs}}$ since $\gamma_d\ll\gamma_{\text{abs}}$, accumulates energy over many pulses through coherent buildup, until saturation at a level set by $\gamma_{\text{abs}}$.

\section{Open-Dirac singularities}

\begin{figure}
    \centering
    \includegraphics[width=1\linewidth]{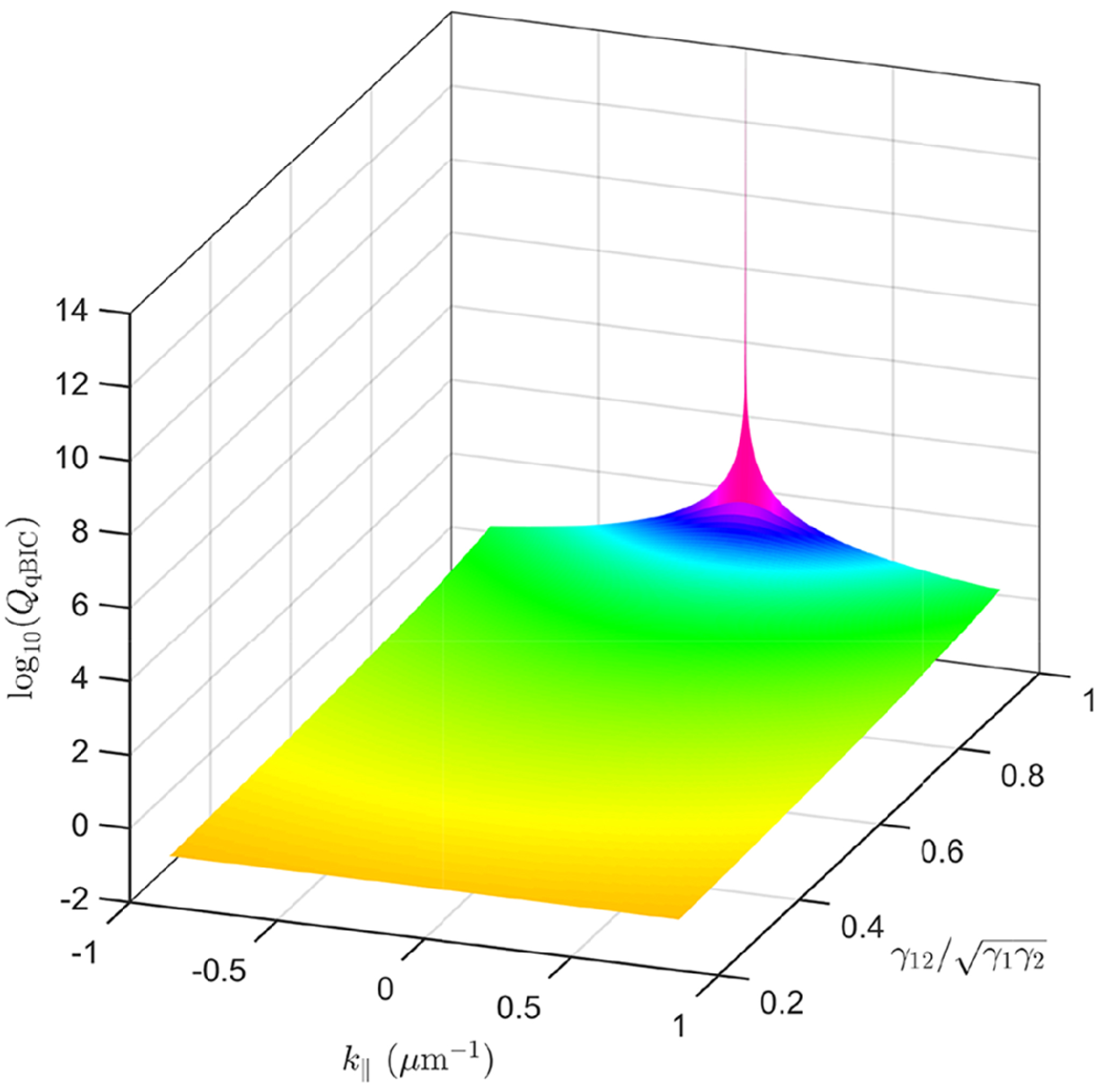}
    \caption{Quality factor of the dark mode as a function of the in plane momentum and ratio $\gamma_{12}/\sqrt{\gamma_1\gamma_2}$ of the original degenerate modes ($\omega_1 = \omega_2$): only when the single radiation channel condition is fulfilled (ratio 1), the dark mode has diverging radiative $Q$-factor because of interference. }
    \label{fig:5}
\end{figure}

\noindent  The two   degenerate physical waves at the crossing $\mathbf{\Gamma}$ point (``closed'' or reference cavity modes, e.g.\ guided modes of the slab without leakage or with idealized boundary conditions) satisfy $\gamma_1(\mathbf{\Gamma})=\gamma_2(\mathbf{\Gamma})$ and $\omega_1(\mathbf{\Gamma})=\omega_2(\mathbf{\Gamma})=\omega_0$. In this limit, the Friedrich--Wintgen condition of Eq.~(\ref{eq:FW}) is automatically fulfilled for any real near-field coupling $\kappa$, so that one of the two supermodes is always a BIC. Indeed, Eqs.~(2)--(5) reduce to
\begin{align}
\omega_b &= \omega_0+\kappa,\\
\omega_d &= \omega_0-\kappa,\\
\forall \kappa\in\mathbb{R},\qquad \kappa'&=0 \Leftrightarrow \text{Eq.~(\ref{eq:FW}) is fulfilled.}\nonumber
\end{align}
\noindent Considering the in-plane symmetry argument of \textbf{Sec. III}, the at-$\mathbf{\Gamma}$ symmetry enforcing destructive interference is preserved and therefore $\kappa'=0$. Only away from $\mathbf{\Gamma}$ is this symmetry lifted, producing a finite phase difference between the radiation channels and thus a non-zero reactive coupling $\kappa'(\mathbf{k})$. In principle, this coupling reaches the supercritical condition $\kappa'=\kappa_{\mathrm{sc}}$ at a finite in-plane momentum $\mathbf{k}_{\mathrm{sc}}\neq\mathbf{\Gamma}$, corresponding to a specific non-normal incidence. In the following, we determine this condition explicitly for Dirac-like dispersion.\\
\indent Dirac cones arise naturally in photonic crystal slabs from symmetry \cite{he2015}. In a 1D grating, a two-wave model is often sufficient, whereas in a 2D PhCS a Dirac cone may emerge from two linearly dispersing transverse electric-like (TE-like) modes coupled to an accidentally degenerate transverse magnetic-like (TM-like) mode, thus forming a three-wave system \cite{huang2011}. In metasurfaces, the TE response more generally involves four coupled waves, while TM excitation often reduces to a two- or three-wave system with opposite sign of $\kappa$.\\
\indent The linear modes form a Dirac cone gapped by a mass term induced by their coupling. This gap is controlled by geometry, especially by out-of-plane inversion asymmetry, for example in an air/PhCS/glass stack. When opened, it produces two symmetry-protected BICs: one in a Dirac branch and one in the slow mode inside the gap, as observed experimentally and numerically \cite{Romano2020,DeTommasi2023,Zito2024}. Here we show that this class of symmetry-protected BICs, although rooted in Dirac physics, is still governed by Friedrich--Wintgen-type interference between modes radiating into the same channel.\\
\indent We first derive the Hamiltonian describing the TE-like Dirac-cone structures relevant to PhCSs and metasurfaces. We then define the \textit{open-Dirac singularity}---in analogy with recent work \cite{Contractor2022}---as the Dirac-gap condition at which supercritical coupling is reached.\\
\indent Let us consider first a two-wave dynamics with group velocity $\pm v_g$. The gapped Dirac Hamiltonian is:
\begin{equation}
H(k) = v_g k_x \sigma_x + \kappa \sigma_z,
\end{equation}
where $\sigma_x$ and $\sigma_z$ are the Pauli matrices, and $k_x = k$ is the in-plane momentum. The representation with off-diagonal $\kappa$ gives a precise  scaling with momentum:
\begin{equation}
\Omega(k) = 
\begin{bmatrix}
\omega_{1} + v_g k & \kappa \\
\kappa & \omega_{2} - v_g k
\end{bmatrix},
\end{equation}
while
\begin{equation}
\Gamma_{\text{rad}} = 
\begin{bmatrix}
\gamma_{1} & \gamma_{12} \\
\gamma_{12} & \gamma_{2}
\end{bmatrix}.
\end{equation}
The two cavity modes of frequencies $\omega_{1,2}(k)$ scale linearly with momentum, while we limit to a uniform radiation loss (without loss of generality) with same previous structure of far-field coupling $\gamma_{12}$. A BIC can emerge into one of the two eigenmodes only by imposing that the rank of the decay matrix is 1. This constraint  is achieved when $\gamma_{12} = \sqrt{\gamma_{1} \gamma_{2}}$. The diagonalization of the non-Hermitian matrix $H(k)$ provides the eigenmodes’ evolution illustrated in \textbf{Figure~\ref{fig:5}}. 
\begin{figure*}
    \centering
    \includegraphics[width=0.8\linewidth]{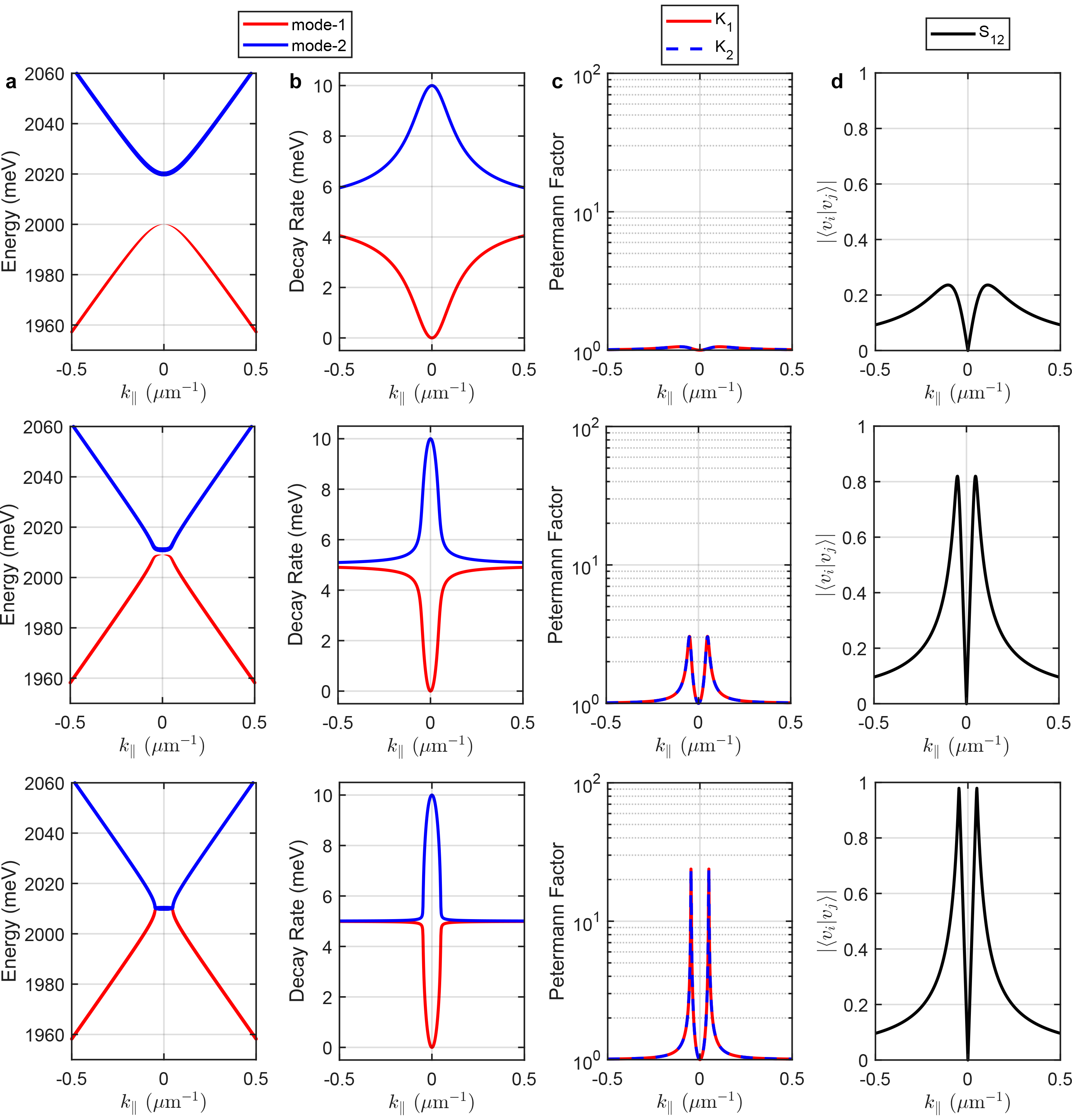}
    \caption{(a) Real part of the eigenmodes (1 and 2, respectively) after diagonalization of the Hamiltonian representing degenerate Dirac modes at $\mathbf{\Gamma}$-point, $\gamma_1=\gamma_2 = 5$ meV, with $v_g=127$meV$\mu$m, as a function of coupling, from top to bottom respectively $\kappa=10, 1.0, 0.1$ meV. The linewidth of the eigenmodes is given by the size of the data points. (b) Associated decay rates. (c) Associated Petermann factor, diverging at EPs, revealing nearly coalescent vectors for reducing gap $2\kappa$. (d) Conventional inner product of right eigenvectors (biorthogonality is verfied, not shown), revealing mutual non-orthogonality. The dark mode EP has radiative loss converging to zero at $\mathbf{\Gamma}$. }
    \label{fig:6}
\end{figure*}

The Q-factor of the negative branch is plotted as a function of $\gamma_{12}$, which underscores that only when a single independent channel is established can a diverging quality factor (BIC) be obtained, which demonstrates the Friedrich-Wintgen origin of the BIC. \\
\indent Let $\gamma_{1}(\mathbf{\Gamma}) = \gamma_{2}(\mathbf{\Gamma}) = \gamma$ and $\omega_{1}(\mathbf{\Gamma}) = \omega_{2}(\mathbf{\Gamma}) = \omega_0$. The coupling strength $\kappa$ can be tuned by the geometry of the system. For TE-like modes, we found that the larger the effective index, the larger the coupling; the lower the vertical asymmetry (i.e., top cladding matching bottom substrate index), the lower the coupling, even approaching zero. However, as can be seen from \textbf{Figure~\ref{fig:6}}, as the Dirac gap $2\kappa \ll \gamma_{12}$, the formation of BIC is accompanied by an EP. Exceptional points are singular spectral degeneracies unique to non-Hermitian systems, at which not only the eigenvalues but also the corresponding eigenvectors coalesce \cite{li2023exceptional, wan2026spin}. The non-Hermitian character of the modes is quantified by the Petermann factor $K_l$ (for the $l$-th mode), as defined in Eq.~S14 (Supporting Information, Sec. I), which accounts for the biorthogonal nature of the left and right eigenvectors and their associated non-orthogonality. This parameter provides a direct measure of excess noise and modal sensitivity in open and lossy photonic systems. Notably, $K_l$ is predicted to diverge at the EP signaling a breakdown of the modal basis and a dramatic amplification of non-Hermitian effects in their vicinity \cite{petermann2003calculated,heiss2012physics,wiersig2023petermann}. Correspondingly, the \textit{right} eigenvectors, $\ket{v_{i,j}} = \ket{R_{i,j}}$, are clearly mutually non-orthogonal as indicated by the evaluated inner products between modes.
 
\indent Now, we will generalize to the typical situation encountered in our 2D PhCS. 
We introduce a third mode \( x_3 \) with quadratic dispersion:
\[
\omega_3(k) = \omega_Q + \beta_Q k^2,
\]
and negligible radiation loss in the TE-polarization channel \( \gamma_Q \approx 0 \).
The full Hamiltonian is:
\[
H =
\begin{bmatrix}
\omega_0 +v_g k + i\gamma & \kappa+i\gamma & \rho \\
\kappa+i\gamma & \omega_0-v_g k + i\gamma & -\rho \\
\rho & -\rho & \omega_Q + \beta_Q k^2 + i\gamma_Q
\end{bmatrix},
\]
where \( \rho \) is the coupling with the slow quadratic mode. This must be antisymmetric for the exchange between $x_1\leftrightarrow x_2$ to reproduce the right curvature at $\mathbf{\Gamma}$, which is in agreement with the results of RCWA and experiments.\\
\indent To avoid redundancy, the associated eigenvalues and decay rates are illustrated as a function of $\kappa$, decreasing from top to bottom panels in \textbf{Supporting Fig.~S2}. Notably, the middle quadratic eigenmode also becomes a quasi-BIC.\\
\indent Since the slow quadratic mode $x_3$ couples symmetrically but with opposite signs to $x_1$ and $x_2$, the interaction Hamiltonian is:
\begin{equation}
H_{\text{int}} = \rho\, x_1^\dagger x_3 - \rho\, x_2^\dagger x_3 + \text{h.c.}
\end{equation}
Transforming again to the bright/dark basis yields:
\begin{align}
H_{\text{int}} &= \rho(x_1^\dagger - x_2^\dagger) x_3 + \text{h.c.} \\
&= \rho\left[\left(\frac{\gamma_1}{\gamma_b} x_b^\dagger + \frac{\gamma_2}{\gamma_b} x_d^\dagger \right) - \left( \frac{\gamma_2}{\gamma_b} x_b^\dagger - \frac{\gamma_1}{\gamma_b} x_d^\dagger \right)\right] x_3 + \text{h.c.} \\
&= \rho\left[\frac{\gamma_1 - \gamma_2}{\gamma_b} x_b^\dagger + \frac{\gamma_1 + \gamma_2}{\gamma_b} x_d^\dagger \right] x_3 + \text{h.c.}\nonumber
\end{align}
Since $\gamma_1 = \gamma_2 =\gamma_b/2 $,
\begin{align}
H_{\text{int}} &= \rho x_d^\dagger x_3 + \text{h.c.}\nonumber
\end{align}
The coupling is purely between the dark mode and the quadratic mode, with zero coupling to the bright mode.
Therefore, rotating the full $H$ in the the basis $\{x_b, x_d, x_3\}$ through $H_{bd} = UHU^\dagger$, with $U = \frac{1}{\sqrt{2}} \begin{pmatrix} 1 && 1 && 0\\ 1&& -1 &&0\\0 && 0 && 1  \\ \end{pmatrix}$, it becomes:
\begin{equation}
H_{bd} =
\begin{bmatrix}
\omega_b + i \gamma_b & \kappa' & 0 \\
\kappa'^* & \omega_d + i \gamma_d & \kappa_{d3} \\
0 & \kappa_{d3}^* & \omega_Q + i \gamma_Q
\end{bmatrix},
\end{equation}
where after proper substitution:
\begin{equation}
H_{bd} =
\begin{bmatrix}
\omega_0 +\kappa+ i \gamma_b & v_g k & 0 \\
v_g k & \omega_0 - \kappa  +i\gamma_d &\rho \\
0 & \rho & \omega_Q + i \gamma_Q
\end{bmatrix}.
\end{equation}
In this basis we find $\kappa' = v_g k$, which is nonzero as expected only away from $\mathbf{\Gamma}$ point, and accounts for any bright--dark coupling away from the ideal at-$\mathbf{\Gamma}$ BIC, whereas $\kappa_{d3} = \rho$ is the direct dark-to-quadratic coupling, and there is no bright-to-quadratic coupling.\\
\indent Assuming weak damping of the dark mode $\gamma_d \ll \gamma_b$ and no direct excitation of $x_3$, the dark mode becomes the energy transfer channel to the quadratic mode.
Using adiabatic elimination of $x_d$:
\begin{equation}
x_d \approx -\frac{\rho  x_3}{\omega_d  +i \gamma_d}.
\end{equation}
Substituting into the equation for $x_3$, yields:
\begin{align}
(\omega_3 + i \gamma_3) x_3 + \rho x_d &= 0 \\
\Rightarrow \left( \omega_3 + i \gamma_Q - \frac{\rho^2}{\omega_d + i \gamma_d} \right) x_3 &= 0.
\end{align}
The effective decay of the slow mode is:
\begin{equation}
\gamma_{\text{eff}} \approx \gamma_Q + \frac{\rho^2 \gamma_d}{\omega_d^2 + \gamma_d^2},
\end{equation}
and maximal energy transfer into $x_3$ (for fixed $\gamma_d$, $\gamma_Q$) occurs when
\begin{equation}
\kappa_{d3,\text{opt}} \sim \sqrt{\gamma_d \gamma_Q} \simeq \gamma_{\text{abs}}.
\end{equation}
This is directly analogous to the earlier bright-mediated optimal condition, but now the energy is routed through the dark mode only, ensuring minimal radiative loss.
This antisymmetric configuration creates an energy-transparent pathway from a nominally inaccessible dark mode into a slow trap. It leverages destructive interference to suppress bright-mode interactions, and the bright reactive coupling to divert input power to the BICs. The system remains effectively subradiant, ideal for trapping, sensing, or nonlinear enhancement.
The supercritical coupling condition for energy transfer from bright to dark mode remains valid in the three-mode case as:
\begin{equation}
\boxed{ \kappa'_{\text{opt}} = \sqrt{ \gamma_b \cdot \gamma_{\text{eff}} } \quad \text{with} \quad \gamma_{\text{eff}} = \gamma_d + \frac{2\rho^2 \gamma_Q}{(\omega_0 - \omega_Q)^2 + \gamma_Q^2} }.
\end{equation}
Since $\gamma_d$ is limited by $\gamma_{\text{abs}}$ as well as $\gamma_Q$, we find again:
\[
\boxed{\kappa'_{\text{opt}} \sim \sqrt{\gamma_b \gamma_d}\simeq \sqrt{\gamma_b \gamma_{\text{abs}}}} 
\]
where \( \gamma_d \simeq\gamma_{\text{abs}} \ll \gamma_b \) and we include nonradiative loss as the lower bound.\\
\indent The gap opened at the Dirac point per construction is:
\[
\Delta_{\text{gap}} = 2(\kappa + \rho),
\]
which becomes $2\kappa$ in a two-wave system. 
The real part of the eigenfrequencies are:
\begin{equation}
\Re[\omega_{\pm}(k)] = \omega_0 \pm \sqrt{\kappa^2 + (v_g k)^2}.
\end{equation}
To have $\kappa'_{\text{opt}}=\kappa_{sc}$ we must consider mode detuning with momentum, at rate determined by their $v_g$, such that $\kappa_{sc} = v_g k_{sc}$. This becomes an equation for a supercritical wavevector, 
\begin{equation}
k_{sc} = \frac{\sqrt{\gamma_b \gamma_{\text{abs}}}}{v_g},
\end{equation}
giving an energy separation 
\begin{eqnarray}\Delta\omega(k_{sc})&=&\Re[\omega_+(k_{sc})-\omega_-(k_{sc})]\\\nonumber &=& 2\sqrt{\kappa^2 + v_g^2k_{sc}^2}\\\nonumber &=& 2\sqrt{\kappa^2 + \gamma_b\gamma_{\text{abs}} }.
\end{eqnarray}  
Substituting realistic values of a typical PhCS used in our previous works and in excellent agreement with RCWA simulations, we set $v_g  = 131.5$ meV$\mu$m, $\gamma_b =$ 10 meV, $\gamma_{\text{abs}}=$ 0.1 meV, for crossing wavelength $\lambda = 620$ nm. 
This yields an angle of incidence that satisfies supercritical coupling $\theta_{\text{sc}} = 0.0428^\circ$ ($k_{\text{sc}} \simeq 8\times10^{-3}\mu\text{m}^{-1}$), for energy separation, say $\kappa = 0.8$ meV, equal to $\Delta\omega = $ 2.56 meV. Realistic values will be validated later in \textbf{Section} \textbf{VIII} through RCWA simulations. \\
\section{Ring of Exceptional Points at the Supercritical Dirac Gap}

\noindent We now determine the exceptional-point condition in the vicinity of the Dirac singularity. 
Neglecting higher-order corrections (see Supporting Information, Sec.~3), the bright/dark Hamiltonian reads
\begin{equation}
H_{bd}=
\begin{pmatrix}
\omega_0+\kappa+i\gamma_b & v_g k\\
v_g k & \omega_0-\kappa
\end{pmatrix}.
\end{equation}

\noindent The complex eigenvalues are
\begin{equation}
\omega_\pm(k)=\omega_0+\frac{i\gamma_b}{2}
\pm \frac{1}{2}\sqrt{4(v_g k)^2+4\kappa^2-\gamma_b^2+4i\gamma_b\kappa}.
\label{eq:EP_main}
\end{equation}
EPs occur when the discriminant vanishes. 
In the small-gap limit $\kappa\ll\gamma_b$, this yields a pair of EPs at
\begin{equation}
k_{\mathrm{EP}}=\frac{\gamma_b}{2v_g}.
\end{equation}

\noindent Since $k=|\mathbf{k}|$, this defines a \emph{ring} of EPs in momentum space. 
This behavior is consistent with previous observations in photonic crystal slabs \cite{zhen2015spaw}, but here it emerges naturally from the bright/dark supermode framework without imposing vanishing radiation of the dark mode or assuming $\kappa=0$. This result is also in agreement with very recent experimental works, where the BIC converges to a ring of EPs \cite{wang2025photoswitchable}.\\
\indent Importantly, the EP condition must be consistent with the supercritical coupling condition derived previously. 
A full treatment including symmetry breaking and loss (Supporting Information, Sec.~3) shows that the Dirac gap value associated with supercritical coupling is not an unrelated fitting parameter, but the same scale that controls the onset of exceptional-point physics. Following the detailed derivations, two conditions must be met: 
\begin{equation}
k_{\mathrm{EP}}\simeq \frac{\gamma_b}{2v_g}, 
\qquad
\Delta_{\mathrm{EP}}=\sqrt{\Gamma_b\Gamma_d}=v_g k_{\mathrm{sc}},
\label{eq:EP_result}
\end{equation}
where $\Delta_{\mathrm{EP}}$ is the semi energy gap  of the Dirac cone and must be equal in size to supercritical coupling separation $v_g k_{\mathrm{sc}}$. This result establishes a key connection: the Dirac gap at which EPs occur coincides with the supercritical coupling condition. In this sense, supercritical coupling is intrinsically EP-correlated: it occurs in the regime where eigenvectors become strongly non-orthogonal and the system approaches the EP ring without necessarily reaching exact coalescence. 
While the EP ring is set by the radiative decay $\gamma_b$, the gap is controlled by total losses $\Gamma_{b,d}$ and therefore depends on absorption.
 As a consequence, supercritical coupling is reached in a regime where the system approaches, but does not fully reach, eigenvector coalescence. 
The Petermann factor therefore remains finite but strongly enhanced, reaching its local 
maximum along the momentum axis at $k_{\mathrm{sc}}$. 
 These results demonstrate that supercritical coupling is intrinsically linked to the topology of the Dirac singularity. \\

\begin{figure*}
    \centering
    \includegraphics[width=0.9\linewidth]{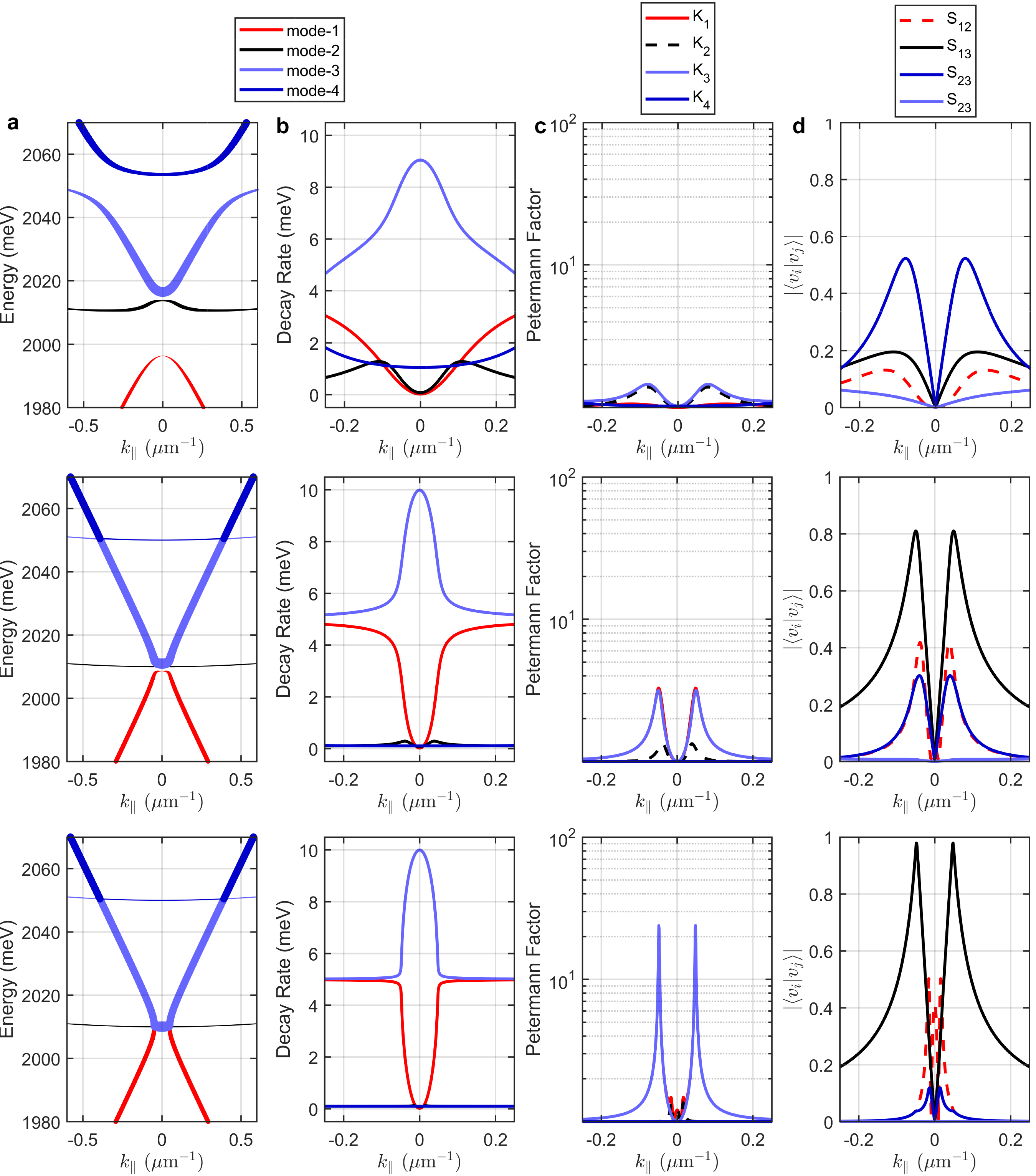}
    \caption{(a) Real part of the eigenmodes (1 = dark and 2 = quadratic $Q$, 3 = quadratic $S$, 4 = bright, respectively) after diagonalization of the 4-wave Hamiltonian, for $\gamma_1=\gamma_2 = 5$ meV, with $v_g=127$meV$\mu$m, as a function of coupling, from top to bottom respectively $\kappa=10, 1.0, 0.1$ meV. The linewidth of the eigenmodes is given by the size of the data points. (b) Associated decay rates. (c) Associated Petermann factor, diverging at EPs, revealing nearly coalescent vectors for reducing gap. (d) Conventional inner product of right eigenvectors, highly non-orthogonal.}
    \label{fig:8}
\end{figure*}
\begin{figure*}
    \centering
    \includegraphics[width=0.9\linewidth]{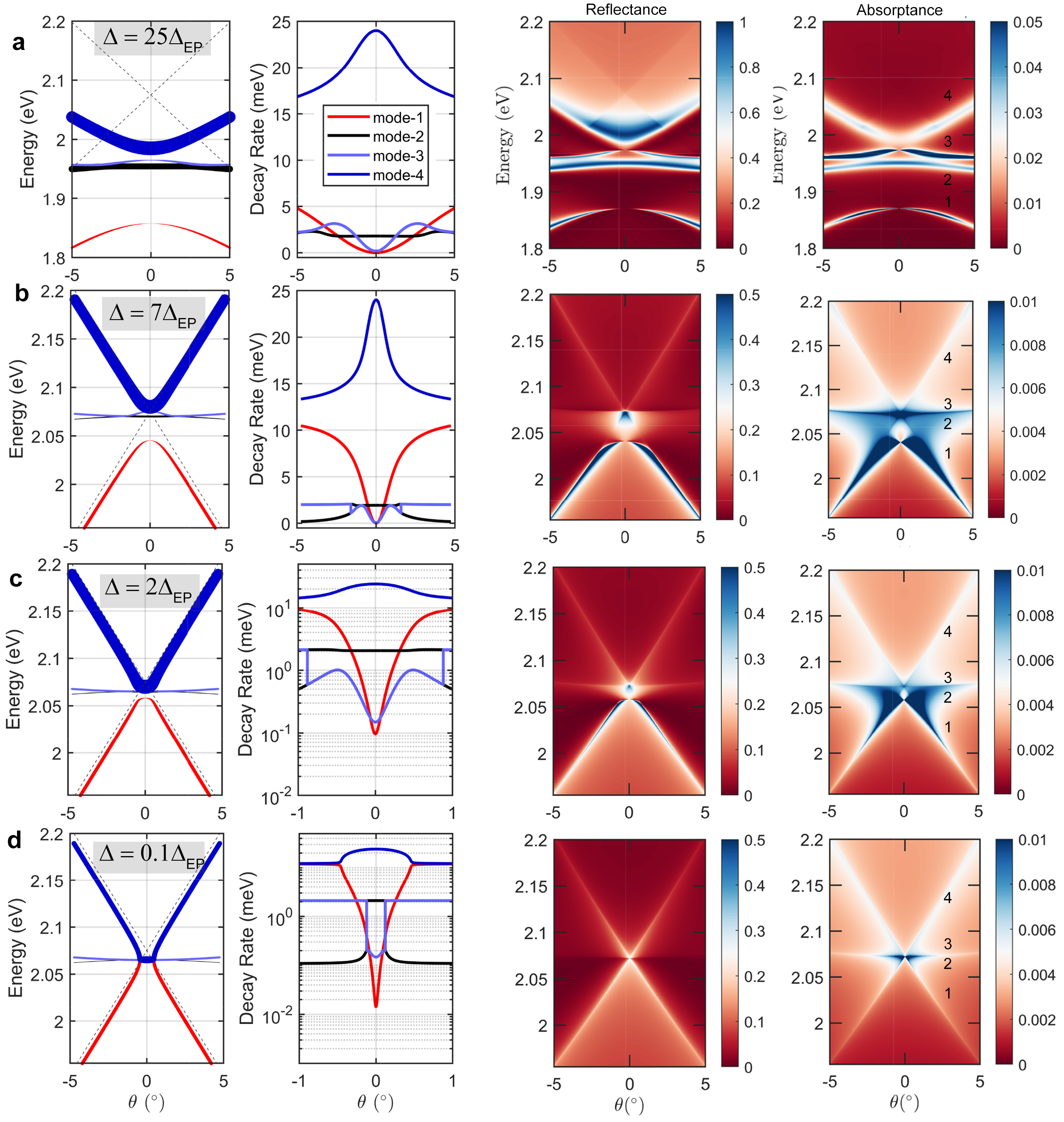}
    \caption{(a) From left to right: Real part of the eigenmodes (1 = dark and 2 = quadratic $Q$, 3 = quadratic $S$, 4 = bright, respectively) after diagonalization of the 4-wave Hamiltonian, for $\gamma_1=\gamma_2 = 12$ meV ($\gamma_b = 24$ meV), $\gamma_Q =2$ meV $\gamma_S = 0.1$ meV, with $v_g=137$meV$\mu$m, at coupling $\kappa = 55$ meV, $\rho = 0.1\kappa$, $h=0.5\kappa$, leading to the indicated gap $\Delta = 25 \Delta_{\text{EP}}$ (the linewidth of the eigenmodes is given by the size of the data points).  Associated decay rates. TE-like Reflectance band diagram obtained through RCWA, in excellent agreement with the model fitted to this geometry.  Corresponding Absorptance band diagram, showig further details. (b)-(d) Same as in (a) but for gap approaching the supercritical gap, respectively $\Delta = 7\Delta_{\text{EP}}$ (b), $\Delta = 2\Delta_{\text{EP}}$ (c), $\Delta = 0.1\Delta_{\text{EP}}$ $\simeq$ 0.25 meV (d).}
    \label{fig:9}
\end{figure*}
\begin{figure*}
    \centering
    \includegraphics[width=1.\linewidth]{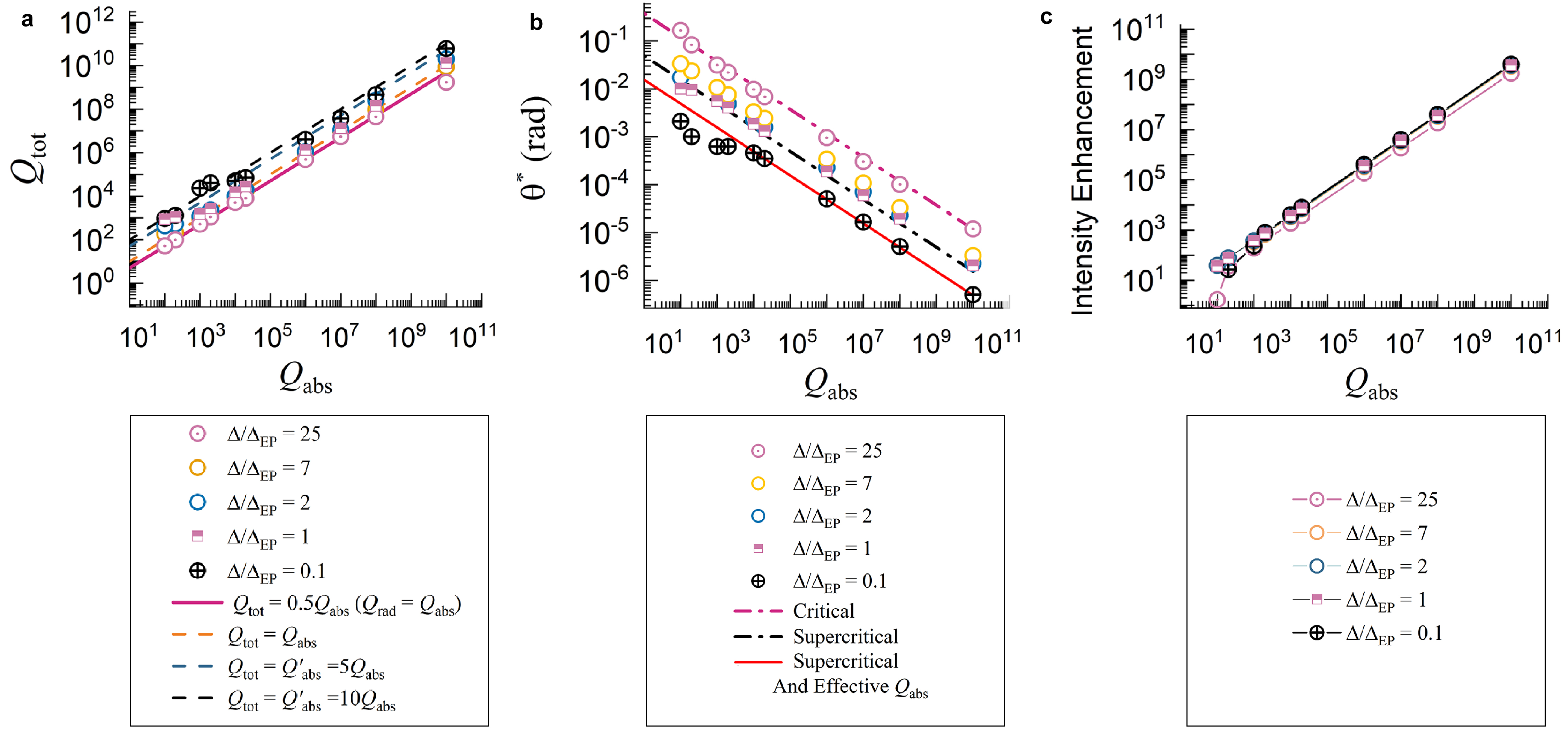}
    \caption{(a) Total quality factor of the quasi-BIC resonance fitted from RCWA simulations as a function of $Q_{\text{abs}}$, included through material loss of the PhCS. The different gap conditions are obtained by changing the radius/lattice period ratio of the PhCS. (b) Angle at which the absorptance and intensity field enhancement result maximum from RCWA simulations as a function of $Q_{\text{abs}}$, for different gap conditions. Away from supercritical coupling regime the behavior is well fitted from critical coupling prediction. Approaching the supercritical gap, only supercritical coupling model predicts the fitted angles. Below the supercritical gap, the data can be interpreted through supercritical coupling model but only admitting an interference-mediated effective absorption. (c) Associated Intensity enhancement.  }
    \label{fig:10}
\end{figure*}
\begin{figure*}
    \centering
    \includegraphics[width=0.8\linewidth]{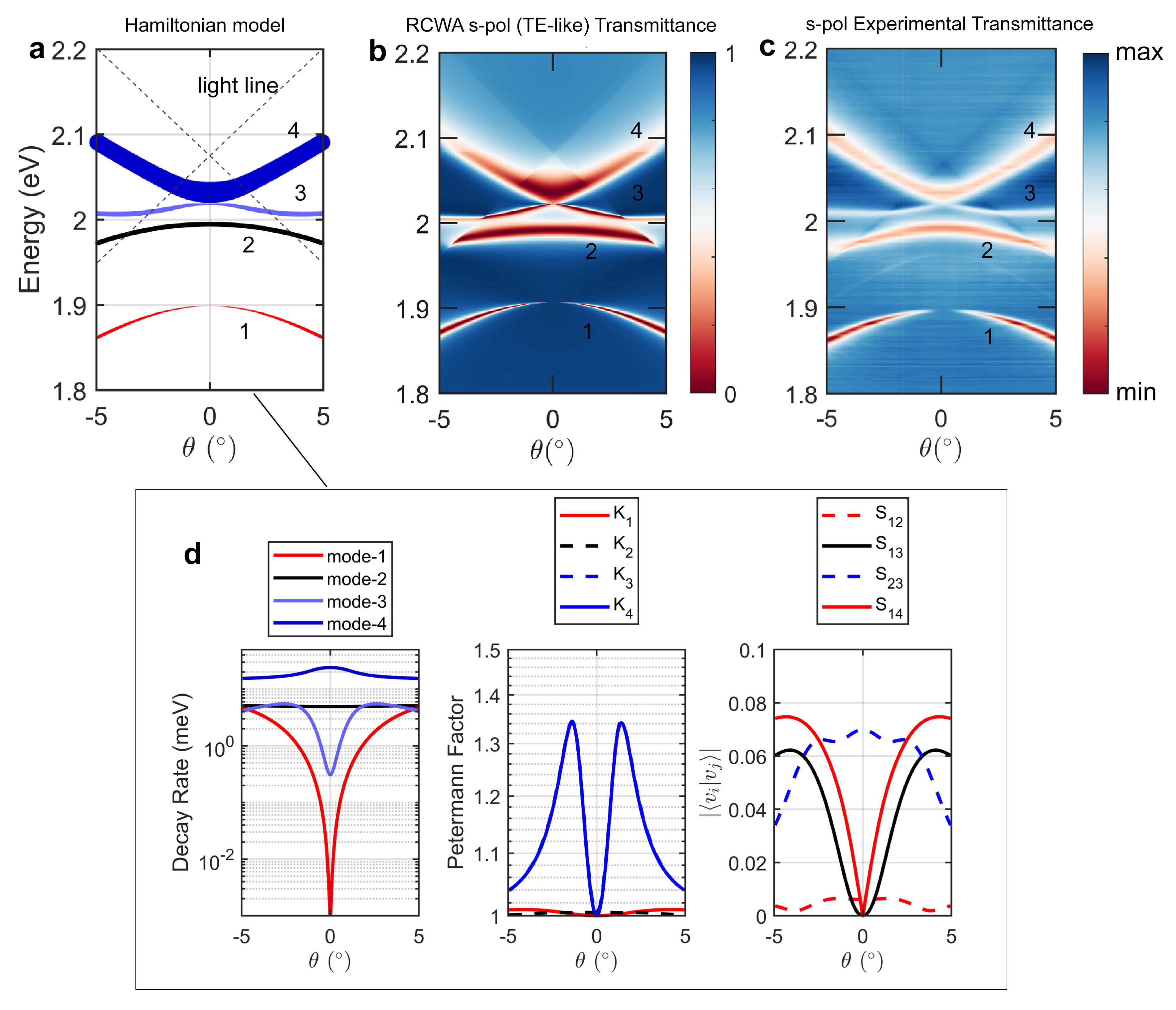}
    \caption{(a) Real part of the eigen energies (1 = dark and 2 = quadratic $Q$, 3 = quadratic $S$, 4 = bright, respectively) after diagonalization of the 4-wave Hamiltonian, for $\gamma_1=\gamma_2 = 12$ meV, with $v_g=137$meV$\mu$m. (b) Comparison with predicted transmittance from RCWA simulations ($a$ = 410 nm, $r$ = 105 nm, $h$= 183 nm). (c) Experimental transmittance of the actually fabricated PhCS showing excellent agreement with both Hamiltniain model and RCWA predictions.  (d) Predicted decay rates, Petermann factors and inner products of right eigenvectors as directly calculated through the Hamiltonian model. Notably, the Hamiltonian analysis operates on a timescale of seconds, whereas accurate RCWA simulations require hours, and experimental fabrication typically spans several days.}
    \label{fig:11old}
\end{figure*}

\section{RCWA validation: beyond critical coupling}

\noindent The TE-like response more generally observed in realistic PhCSs and metasurfaces is described by a four-wave system. We therefore introduce a fourth mode of amplitude \(x_4\), dispersion
\begin{equation}
\omega_4(k)=\omega_S+\beta_S k^2,
\end{equation}
and loss \(\gamma_S\). As for the first quadratic mode, this second slow mode must couple with opposite sign to the positive- and negative-slope Dirac branches in the original basis. This choice, consistently supported by RCWA as a function of geometry, reproduces the observed mode swapping, the correct repulsion between quadratic and Dirac branches as their relative detuning changes, and the emergence of EPs between quadratic modes under suitable conditions. 
The resulting effective Hamiltonian is
\begin{widetext}
\begin{equation}
H =
\begin{bmatrix}
\omega_0 + v_g k & \kappa & \rho & h \\
\kappa & \omega_0 - v_gk & -\rho & -h \\
\rho & -\rho & \omega_Q + \beta_Q k^2 & 0 \\
h & -h & 0 & \omega_S + \beta_S k^2
\end{bmatrix}
+i\begin{bmatrix}
\gamma & \gamma  & 0 & 0 \\
\gamma & \gamma  & 0 & 0 \\
0 & 0 & \gamma_Q & 0\\
0 & 0 & 0 & \gamma_S 
\end{bmatrix}.
\end{equation}
\end{widetext}
\textbf{Figure~\ref{fig:8}} shows the eigenmode evolution of this four-wave system as a function of \(\kappa\). As in the three-wave case, EP formation occurs for small \(\kappa\), i.e. when the gap approaches the supercritical regime. Importantly, although the additional waves reshape the spectrum, the value of \(\kappa_{\mathrm{sc}}\) remains nearly unchanged, confirming that the supercritical condition is robust and controlled primarily by the bright/dark sector. We therefore use this full Hamiltonian to validate the theory against RCWA.\\
\indent We performed RCWA simulations for a silicon nitride PhCS consisting of a two-dimensional square lattice of air holes, with period \(a=410\) nm, slab thickness \(h=183\) nm, and hole radius in the range \(r=(105,175)\) nm, using the measured refractive-index dispersion reported in Ref.~\cite{Zito2024}. These values correspond to fabricated and experimentally tested samples, some of which are shown in \textbf{Supplementary Fig. S3}. The mode coupling is tuned by varying the ratio \(r/a\): increasing \(r/a\) lowers the effective index and therefore reduces the gap, while increasing slab thickness increases the effective index and enlarges the coupling. We also varied the material loss by introducing an imaginary part in the refractive index, \(n=n_R+i n_I\), with \(n_I>0\) under the convention \(e^{i(kz+\omega t)}\). The corresponding intrinsic absorption quality factor is \(Q_{\text{abs}}=n_R/2n_I\), while \(n_R(\omega)\) was kept fixed. We explored a broad range of losses, from nearly ideal transparent systems (\(n_I\sim10^{-10}\)) to moderately lossy ones (\(n_I\sim10^{-2}\)). In particular, \(n_I\in(10^{-6},10^{-4})\) corresponds to high-quality SiN, whereas larger values are representative of doped or hybrid platforms such as SiN loaded with emitters or absorbers, or slabs made of more dissipative materials such as transition-metal dichalcogenides or perovskites away from resonance. Since the absorption quality factor of the photonic modes depends on field distribution and modal overlap, we vary the imaginary refractive index of the whole PhCS region for simplicity and direct comparison.\\
\indent The results for a representative value \(n_I=10^{-4}\) are shown in \textbf{Figure~\ref{fig:9}}, which compares the fitted eigenmode dispersion and decay spectra with RCWA-calculated reflectance and absorptance maps of the TE-like modes along the \(\mathbf{\Gamma X}\) direction. From top to bottom, the coupling is systematically reduced, corresponding to \(\Delta=(25,7,2,0.1)\Delta_{\text{EP}}\). The agreement between the Hamiltonian and RCWA is very good over the full range, including the narrowing of the gap and the approach to the supercritical regime. As the gap decreases, eigenmodes 2 and 3 become nearly indistinguishable, while the EPs become increasingly visible. Equivalent RCWA calculations along \(\mathbf{\Gamma M}\) reveal the same topology, confirming that the EPs form a rotationally symmetric ring in momentum space.\\
\indent The full sweep over \(n_I\) is summarized in \textbf{Figure~\ref{fig:10}}, where we extract the angle
\[
\theta^\star=\arcsin\!\left(\frac{k_\parallel^\star}{k_0}\right),
\]
with \(k_0=2\pi/\lambda=\omega/c\), at which the maximum field enhancement occurs. Operationally, \(\theta^\star\) is determined from the maximum absorptance peak,
\(
A_{\max}=A(\omega,k_\parallel^\star)=1-T(\omega,k_\parallel^\star)-R(\omega,k_\parallel^\star).
\)
We then fit the corresponding total quality factor \(Q_{\text{tot}}\) from the Lorentzian linewidth of the absorptance peak at \(\theta^\star\).
\textbf{Figure~\ref{fig:10}a} shows \(Q_{\text{tot}}\) as a function of \(Q_{\text{abs}}\) for different values of \(\Delta/\Delta_{\text{EP}}\). For large gap, the results follow the conventional critical-coupling condition \(Q_{\text{rad}}=Q_{\text{abs}}\), as expected.
 A qualitatively different behavior appears as the gap approaches \(\Delta_{\text{EP}}\). The fitted quasi-BIC linewidth becomes not only much narrower than expected from radiation leakage, but in fact even narrower than what absorption alone would impose, leading to values of \(Q_{\text{tot}}\) up to an order of magnitude larger than the nominal \(Q_{\text{abs}}\). This effect was systematic, and also related to the absorptance pattern of \textbf{Fig.~\ref{fig:9}d}, where dissipation is at least partly shifted toward the quadratic branches.\\
\indent The central result is shown in \textbf{Figure~\ref{fig:10}b}.  Under critical coupling regime, the optimal angle must satisfy
\[
Q_{\text{rad}}(\theta^\star)=Q_{\text{abs}}(\theta^\star),
\]
with \(Q_{\text{abs}}=\alpha n_I^{-1}\) and \(Q_{\text{rad}}\sim Q_0/(\sin\theta)^{2m}\) \cite{Koshelev2018}. This leads to
\begin{equation}
\theta^\star=\arcsin\left[\left(\frac{Q_0 n_I}{\alpha}\right)^{1/2m}\right].
\label{eq:Critical}
\end{equation}
The typical case is \(m=1\), while merging-BIC scenarios may yield \(m>1\), which implies larger \(\theta^\star\) because the radiative quality factor becomes a slower function of \(k-k_{\text{BIC}}\). For \(\Delta/\Delta_{\text{EP}}=25\), the numerical data follow the critical-coupling prediction very well, and the optimal angle increases with loss exactly as expected from Eq.~\eqref{eq:Critical}. In stark contrast, as \(\Delta/\Delta_{\text{EP}}\to1\) and below, the values of \(\theta^\star\) extracted from RCWA depart strongly from the critical-coupling curve. In this regime, the condition \(Q_{\text{rad}}=Q_{\text{abs}}\) is no longer the correct criterion for maximizing the dark-mode enhancement. Instead, the data are captured by the supercritical relation
\begin{equation}
\theta_{\text{sc}}^\star
=
\arcsin(k_{\text{sc}}/k_0)
=
\arcsin\left(\frac{2\pi\sqrt{\gamma_b\gamma_{\text{abs}}}}{\lambda v_g}\right).
\end{equation}
Using RCWA-extracted parameters (\(v_g=137\) meV\(\mu\)m, \(\gamma_b=24\) meV, \(\gamma_{\text{abs}}=2/(\omega Q_{\text{abs}})\)), this expression quantitatively reproduces the black dashed curve in \textbf{Figure~\ref{fig:10}b}. For example, at \(Q_{\text{abs}}=10^4\), one has \(\gamma_{\text{abs}}=0.2073\) meV, and the quasi-BIC appears at \(\lambda=598\) nm, corresponding to \(k_0=10.47~\mu\mathrm{m}^{-1}\).\\
\indent The red curve in \textbf{Figure~\ref{fig:10}b} is instead obtained by applying the same supercritical model with an effective absorption \(Q'_{\text{abs}}=10Q_{\text{abs}}\), extracted from the measured \(Q_{\text{tot}}\) under the assumption \(Q_{\text{rad}}\gg Q_{\text{abs}}\). In parallel, \textbf{Figure~\ref{fig:10}c} shows that the field enhancement increases beyond the limit expected from the nominal material absorption.\\
\indent Both observations strongly indicate that as the gap is reduced even beyond the open-Dirac singularity, a more complex scenario establishes that gives rise to an absorption damping. This will be analytically examined in the next section,  justifying the effective absorption used above (\(Q'_{\text{abs}}\)) to fit the data, and validated through inspecting the field distributions extracted from RCWA simulations. Notably, reproducing the same \(\theta^\star\) within a standard critical-coupling model would require an effective \(Q'_{\text{abs}}\sim 10^3 Q_{\text{abs}}\), which is two orders of magnitude beyond the RCWA-extracted values and therefore considered highly unrealistic.\\
\indent Finally, \textbf{Figure~\ref{fig:11old}} shows a representative comparison between the Hamiltonian model, RCWA, and experimentally measured transmittance of a real PhCS. The diffraction light line is only slightly shifted relative to theory because of small differences in substrate dispersion. We also note that the third eigenmode tends toward a symmetry-protected BIC with enhanced field concentration and nontrivial topological character. A similar behavior was already visible in experiment/simulation comparisons previously reported in Ref.~\cite{Zito2024}, for the three-wave case. Additional comparison between RCWA and experimental results is further shown in \textbf{Supplementary Fig. 3} as a function of the hole radius of the PhCS unit cell.\\
\indent Moreover, an extension of the current four-wave model including also exciton dispersion is reported in Ref. \cite{miranda2026}. The extended model allows engineering strong coupling and polariton branch dispersion in transition metal dichalcogenide monolayers coupled with PhCSs and is found in excellent agreement with  thorough experimental characterization of fabricated samples.

\section{Non-orthogonality-induced absorption interference}

\noindent 
The agreement between RCWA and the generalized Hamiltonian demonstrates that the quasi-BIC operates in a genuinely non-Hermitian regime. As the Dirac gap approaches $\Delta_{\mathrm{EP}}$, the eigenvectors become strongly non-orthogonal. While the redistribution of radiative loss between coupled modes is well established, the corresponding behavior of absorptive decay channels has received far less attention.
The imaginary part of each eigenfrequency can be decomposed as
\[
\gamma_j = \gamma^{\mathrm{eff}}_{j,\mathrm{rad}} + \gamma^{\mathrm{eff}}_{j,\mathrm{abs}},
\]
where the effective absorption is
\begin{equation}
\gamma^{\mathrm{eff}}_{j,\mathrm{abs}} = 
\langle L_j | \Gamma_{\mathrm{abs}} | R_j \rangle.
\label{eq:gamma_abs_eff}
\end{equation}
This expression shows that absorption is not an intrinsic scalar property of a mode, but depends on the biorthogonal structure of the eigenvectors. Near exceptional points, where eigenvectors become nearly parallel, cross terms are strongly enhanced and enable redistribution of dissipation between modes.
Material absorption originates from the absorbed-power functional
\begin{equation}
P_{\mathrm{abs}}[\mathbf{E}]
=
\frac{\omega}{2}
\int_V
\mathrm{Im}\,\varepsilon(\mathbf{r},\omega)
|\mathbf{E}(\mathbf{r})|^2\,d^3r .
\end{equation}
Projecting this functional onto a modal basis yields the absorptive overlap matrix
\begin{equation}
\mathcal{P}_{\mathrm{abs}}^{mn}
=
\frac{\omega}{2}
\int_V
\mathrm{Im}\,\varepsilon(\mathbf{r},\omega)
\,\mathbf{E}_m^*(\mathbf{r})\cdot\mathbf{E}_n(\mathbf{r})\,d^3r .
\label{eq:Pabs_projection}
\end{equation}
The corresponding normalized absorptive decay matrix is then
\begin{equation}
\Gamma_{\mathrm{abs}}^{mn}
=
\frac{\mathcal{P}_{\mathrm{abs}}^{mn}}
{\sqrt{U_m U_n}},
\label{eq:Gammaabs_projection}
\end{equation}
where
\begin{equation}
U_m=
\frac{1}{4}
\int_V
\left[
\varepsilon_0\,\mathrm{Re}\,\varepsilon(\mathbf{r},\omega)\,
|\mathbf{E}_m|^2
+
\mu_0|\mathbf{H}_m|^2
\right]d^3r
\end{equation}
is the stored electromagnetic energy of mode \(m\), evaluated over the full computational domain.
Even for spatially uniform $\mathrm{Im}\,\varepsilon$, the matrix $\Gamma_{\mathrm{abs}}$ is generally non-diagonal due to modal overlap, enabling interference between absorptive decay channels.
Near the open-Dirac singularity, RCWA reveals extreme non-orthogonality, e.g., from Fig.~\ref{fig:8}d for closing gap, we may find
\[
|\langle R_1|R_3\rangle| \simeq 0.98,
\]
so that off-diagonal absorption terms can become comparable to the diagonal ones. Under these conditions, even in the eigenbasis of the effective Hamiltonian, the absorption operator remains non-diagonal:
\begin{equation}
\Gamma_{\text{abs}} = 
\begin{pmatrix}
\zeta_{1} & \zeta_{12} \\
\zeta_{12}^{\star} & \zeta_{2}
\end{pmatrix}.\label{eq:Gamma_zeta}
\end{equation}
For strong overlap ($\zeta_{12} \sim \sqrt {\zeta_{1}\zeta_{2}}$), the absorption matrix approaches a near rank-1 structure, yielding strongly asymmetric effective decay rates in the absorption operator eigenbasis:
\[
\zeta^{\mathrm{eff}}_b \gg \zeta^{\mathrm{eff}}_d,
\]
i.e., one mode becomes loss-bright, while the other becomes absorption-dark.\\
\indent In Supporting Information, Sec.~4, we quantitatively address this possibility considering it within the four-mode Hamiltonian, showing that absorption interference could actually account, at least in principle, for the increased effective absorption \((Q'_{\text{abs}}\)) observed and used for data fitting in \textbf{Fig.~\ref{fig:10}}. In the next section, the analysis will involve the actual field distributions calculated through RCWA simulations, corresponding to the real eigenvectors of the system.

\section{RCWA Results: field delocalization and absorption interference}
\begin{figure*}[t!]
    \centering
    \includegraphics[width=1\linewidth]{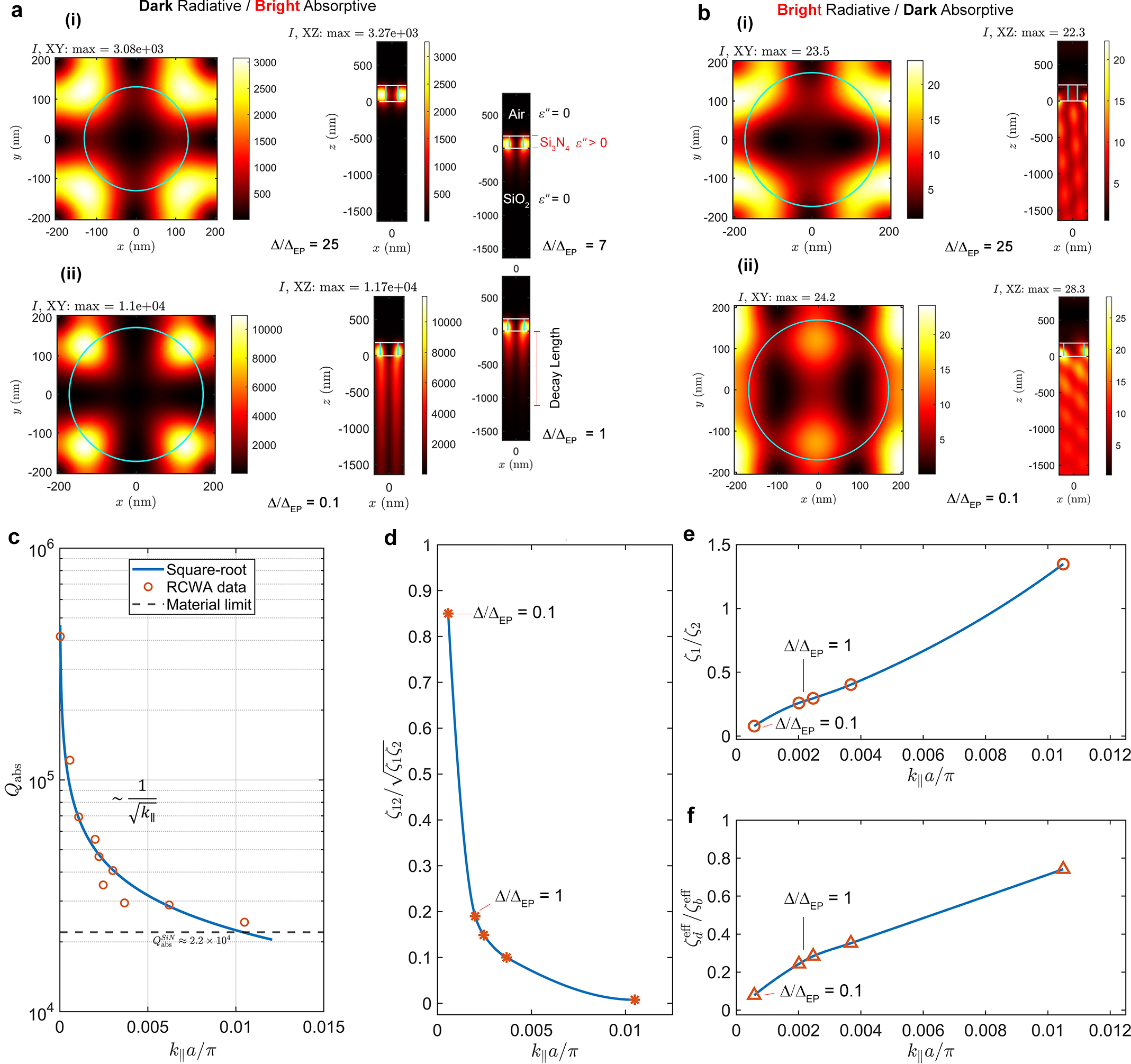}
    \caption{(a) Spatial intensity distributions of the quasi-BIC in the PhCS unit cell exhibiting dark radiative and bright absorptive character, shown in the $xy$ plane (left) and corresponding $xz$ cross-sections (right), for decreasing gap values $\Delta/\Delta_{\mathrm{EP}} = 25, 0.1$ (i, ii). As the system approaches the supercritical regime, the field progressively delocalizes into the low-loss surrounding medium while maintaining stronger and stronger intensity. 
(b) Complementary bright radiative mode, with propagating character evolving towards a field configuration with reduced absorption losses driven by delocalization in the substrate but also in the central cylindrical void of the slab. 
(c) Extracted absorption quality factor $Q_{\mathrm{abs}}$ of the quasi-BIC as a function of normalized in-plane momentum $k_{\parallel}a/\pi$. RCWA data (symbols) follow a square-root scaling (solid line), $ \sim 1/\sqrt{k_{\parallel}}$, consistent with the divergence of the evanescent penetration depth near the light line. The dashed line indicates the intrinsic material absorption limit. 
(d) Normalized absorptive cross-coupling coefficient $\zeta_{12}/\sqrt{\zeta_{11}\zeta_{22}}$ as a function of $k_{\parallel}a/\pi$, revealing the progressive opening of a coherent absorption channel, approaching a near rank-1 condition. (e) Corresponding redistribution of effective absorption between modes, showing the emergence of a loss-bright and loss-dark pair. Top eigenvalues ($\zeta_{1}, \zeta_{2}$)  are from the effective Hamiltonian eigenvectors, while in (f) eigenvalues ($\zeta_{b}^{\mathrm{eff}}, \zeta_{d}^{\mathrm{eff}}$) are from the diagonal loss representation. Markers indicate RCWA-extracted values at representative gap conditions. Blue solid lines interpolating the points are a guide to the eye.
These results demonstrate that modal non-orthogonality provides an absorption-channel analogue of shaping the field overlap with lossless and lossy regions of the resonator.
}
    \label{fig:10Rev}
\end{figure*}
\noindent
We rigorously investigate the role of absorption interference directly  analyzing the RCWA-retrieved  field distributions. This allows us to elucidate why, near the supercritical Dirac condition, the quasi-BIC exhibits a total quality factor $Q_{\mathrm{tot}}$ exceeding the nominal absorption-limited value $Q_{\mathrm{abs}}$ by up to one order of magnitude, and to assess the extent to which this enhancement originates from interference within the absorption channel.
\indent 
\textbf{Delocalization.} Figure~\textbf{\ref{fig:10Rev}a,b} shows the spatial field distributions of the two dominant eigenmodes. A central feature emerging from these maps is the progressive delocalization of the quasi-BIC into the surrounding substrate. The mode undergoes a pronounced reorganization, with an evanescent tail that extends deeply outside the slab. The fitted intensity decay lengths in SiO$_2$ are $160, 180,264,337,1157$ nm (for decreasing gap). The pronounced increase from 160 to 1160 nm is equivalent to the range $(0.27\lambda, 1.93\lambda)$, with $\lambda \simeq 600$ nm. It is wroth noting that the lowest intensity enhancement of the most delocalized field (at its farthest decay point) is $\sim2000$, which is the same order of maximum enhancement deep inside the slab of the less delocalized mode $\sim3374$. This represents an extraordinary increase of spatial envelope over extended regions. This delocalization suppresses the overlap with absorptive media, reinforcing the reduction of $\gamma_{\mathrm{abs}}$. The behavior is governed by the evanescent decay constant
\begin{equation}
k_z = \sqrt{k_\parallel^2 - k_0^2 n^2}.
\end{equation}
As the light line is approached (\(k_\parallel \to k_{\mathrm{light}} = k_0 n\)), one has
\[
k_z \to 0, \qquad \delta \sim \frac{1}{k_z} \to \infty,
\]
leading to a divergence of the penetration depth.
As a direct consequence, the modal overlap with the lossy region (SiN slab) is strongly reduced. The effective absorption is therefore governed by the fraction of field energy remaining inside the absorbing medium, yielding the scaling
\begin{equation}
\gamma_{\mathrm{abs}}^{\mathrm{eff}} \propto k_z,
\qquad
Q_{\mathrm{abs}}^{\mathrm{eff}} \propto \frac{1}{k_z}.
\end{equation}
Expanding close to the light line by writing \(k_\parallel = k_{\mathrm{light}}+\Delta k\), with \(\Delta k \ll k_{\mathrm{light}}\), gives
\begin{equation}
k_z
=
\sqrt{(k_{\mathrm{light}}+\Delta k)^2-k_{\mathrm{light}}^2}
\;\simeq\;
\sqrt{2k_{\mathrm{light}}\Delta k},
\end{equation}
hence
\begin{equation}
Q_{\mathrm{abs}}^{\mathrm{eff}} \propto \frac{1}{\sqrt{k_\parallel-k_{\mathrm{light}}}}.
\end{equation}
This trend is quantitatively confirmed in Fig.~\textbf{\ref{fig:10Rev}c}, where $Q_{\mathrm{abs}}$ exhibits a clear square-root  divergence (solid blue line). \\
\indent The observed delocalization of the quasi-BIC mode can be rigorously interpreted as arising from its hybridization with an extended Bloch wave of the periodic lattice. In this picture, the resonant state acquires the spatial structure of a Bloch mode, whose field can be expressed as $E(\mathbf{r}) = u(\mathbf{r}) e^{i \mathbf{k}_{\parallel} \cdot \mathbf{r}}$, where $u(\mathbf{r})$ retains the lattice periodicity. Crucially, in the present regime the in-plane wavevector approaches the center of the Brillouin zone, $\mathbf{k}_{\parallel} \to \mathbf{0}$ (the light line actually intersects the $\mathbf{\Gamma}$-point), such that the phase factor reduces to the trivial identity
\(
e^{i \mathbf{k}_{\parallel} \cdot \mathbf{r}} = 1.
\)
This seemingly simple limit has profound physical consequences: the Bloch wave no longer carries a spatially varying phase, and the electromagnetic field becomes globally phase-locked across the entire structure. As a result, all unit cells oscillate coherently, and the quasi-BIC transitions from a localized resonance into a collective, aperture-spanning mode. Finally, we emphasize that this regime closely parallels the behavior reported in recent experimental studies \cite{chua2014,Contractor2022,ma2023}. \\

\indent \textbf{Absorption interference.} At this stage, we quantified also the absorptive cross-coupling,  directly evaluated from the overlap integral over the lossy SiN region and surrounding lossless regions forming the full simulation volume $V$,
using Eqs.~\eqref{eq:Gammaabs_projection} and \eqref{eq:Gamma_zeta}.
Figure~\textbf{\ref{fig:10Rev}d} shows the evolution of the figure of merit 
\begin{equation}C=|\zeta_{12}|/\sqrt {\zeta_{1}\zeta_{2}},
\end{equation}
a metric ranging between 0 and 1 that quantifies the opening of the coherent channel between the modes ( $C = 1 \to$ rank-1 absorption matrix condition). The data obtained are reported as a function of the in-plane wavevector (incident angle) where maximal field enhancement occurs, corresponding to decreasing values of $\Delta/\Delta_{\text{EP}}$ from 25 to 0.1.  Approaching the EP point, the absorptive structure in the effective Hamiltonian eigenbasis becomes (transposing rates in units of meV):
\begin{equation}
\Gamma_{\text{abs}} =
10^{-2}
\begin{pmatrix}
1.708 & 0.390 - 1.495\,i \\
0.390 + 1.495\,i & 1.916
\end{pmatrix} \,\mathrm{meV}.
\end{equation}
The diagonal entries $\zeta_{1,2}$, respectively, correspond to the bright-radiative mode and to the dark radiative quasi-BIC. The ratio $\zeta_2/\zeta_1 = 1.12$ corresponds perfectly to the retrieved ratio $Q^{(1)}_{\text{abs}}/Q^{(2)}_{\text{abs}}=1.12$.\\
\indent 
Remarkably, we find that as the modes become highly non-orthogonal $C$ $\simeq 0.854$, revealing the coherent absorption interference directly from the vector field distributions. Moving to the eigenbasis of the absorption operator, the diagonalization leads to the highly asymmetric eigenvalues
\begin{equation}
\zeta^{\text{eff}}_{b} \simeq 3.305\times 10^{-2}\ \mathrm{meV} \gg
  \zeta^{\text{eff}}_{d} \simeq 3.182\times 10^{-3}\ \mathrm{meV},
\end{equation}
with normalized eigenvector approximately
\begin{equation}
\mathbf{v}_{\pm} \approx
\begin{pmatrix}
0.393 \\
\pm0.233 \pm 0.892\,i
\end{pmatrix},
\end{equation}
with corresponding modal weights:
\begin{equation}
|v_{+,1}|^2 \approx 0.154,
\qquad
|v_{+,2}|^2 \approx 0.846.
\end{equation}
Therefore, the absorption-bright eigenmode is composed of approximately
\(85\%\) mode 2 (quasi-BIC and) \(15\%\) mode 1.\\

\indent \textbf{Exchange of the Bright and Dark Character.} This mechanism suggests an absorption-channel analogue of the Friedrich--Wintgen interference where the rank-1 condition drives the emergence of the BIC. In the FW mechanism,   the radiative interference  reconfigures microscopically the mode vector distribution in a vortex-like array that has zero coupling with far field radiation, thereby leading to a zero radiation decay. Intrinsic dissipation in general cannot be regarded as a fixed material parameter since the modal overlap with the lossy regions determines the actual contributions to absorption loss. The point here is that the non-orthogonality allows a coherent redistribution among eigenmodes through the shared absorptive channel, reshaping the corresponding modal structures. The absorption interference reconfigures microscopically the vector fields such that the modal overlap with lossy regions is increased or reduced, enforcing or decreasing the photonic interactions with the material regions through non-Hermitian coupling. The mode experiencing the largest absorptive decay corresponds to the dark-radiative quasi-BIC itself (eigenvalue $\zeta_2$) as shown in \textbf{Fig.~\ref{fig:10Rev}e}, reflecting its strongly enhanced field amplitude, but with shrunk effective mode volume driven by the non-Hermitian coupling. In other words, while coherent radiative interference leads to a dark-radiative mode, the analogue absorptive mixing  flows in the opposite direction leading to a brighter absorptive mode (more concentrated in the absorbing medium when compared with the the coupled partner). In contrast, the bright radiative mode reduces the absorption loss by increasing its modal energy fraction in the void (air regions) of the slab (\textbf{Fig.~\ref{fig:10Rev}b}), with eigenvalues $\zeta_1$ ($\zeta_{d}^{\mathrm{eff}}$ in the absorption diagonal basis) decreasing one order of magnitude more than the coupled $\zeta_2$ ($\zeta_{b}^{\mathrm{eff}}$),  as shown in \textbf{Fig.~\ref{fig:10Rev}e,f}. \\
\indent Although it is well established that the effective absorptive contribution of a mode depends on its field distribution \cite{kristensen2014modes}, no practical strategies or engineering mechanisms have been identified to explicitly harness and control the absorption channel itself in the context of the quasi BICs. Here instead, it emerges as a tangible mechanism, defining a regime in which intrinsic nonradiative losses can be actively reshaped, introducing a new parameter to manipulate light--matter interactions in photonic platforms, of potential interest where material dissipation fundamentally limits performance.

\section{Generalized Bound States in the Continuum}

\noindent
The simulation results discussed above reveal that, in our specific open-Dirac configuration, two distinct counteracting mechanisms govern intrinsic losses:
(i) absorption interference; 
and (ii) field delocalization.
When combined, field delocalization dominates and drives the system toward a limiting state, with overall reduced electromagnetic energy density fraction stored withing the absorbing medium. This leads to the observed scaling of \(Q_{\mathrm{abs}}^{\mathrm{eff}}\), so that its effective absorption rate is strongly reduced of over an order of magnitude.\\
\indent
This diverging absorption quality factor motivates us to conjecture the possibility of achieving a \emph{generalized bound state in the continuum} (gBIC), defined as a state satisfying
\begin{equation}
\mathrm{Im}\,\tilde{\omega} =
\langle L | \Gamma_{\mathrm{rad}} + \Gamma_{\mathrm{abs}} | R \rangle \to 0.
\end{equation}
In this extreme limit, the sum of radiative and absorptive decay channels would be ideally suppressed through coherent interference and/or spatial redistribution over absorbing and lossless domains of the resonators, leading to a simultaneous divergence of the total quality factor.\\
\indent However, in the more general scenario, we must consider the possibility that the sum may approach a zero, but not necessarily each of the individual loss components independently.
In the passive systems considered here, this condition appears to be approached only because of field delocalization, but opening the fundamental question of whether a state with virtually infinite total quality factor, hence unlimited intracavity field enhancement, can be realized. This perspective naturally points toward non-Hermitian platforms with engineered gain--loss distributions, where the sign-indefinite nature of $\varepsilon''(\mathbf{r})$ enables compensation of intrinsic dissipation. In particular, parity--time ($\mathcal{PT}$)-symmetric or more generally balanced gain--loss configurations provide a viable route.\\
\indent The positivity of the material loss, encoded in \(\varepsilon''(\mathbf{r}) \ge 0\), imposes a strict lower bound on the nonradiative decay rate. Formally, the non-Hermitian functional
\begin{equation}
\mathcal{L}_{\mathrm{nr}}[\mathbf{E}]
=
\int dV\, \varepsilon''(\mathbf{r})\,|\mathbf{E}(\mathbf{r})|^2
\end{equation}
remains strictly positive, so that \(\gamma_{\mathrm{nr}}>0\) and a perfectly non-decaying eigenstate cannot exist.\\
\indent
Therefore, we consider systems where the imaginary part of the permittivity is sign-indefinite, $\varepsilon''(\mathbf{r})$, i.e., in the presence of spatially distributed gain and loss. Under these conditions, the nonradiative dissipation functional can be exactly compensated through the balance
\begin{equation}
\int_{V_{\mathrm{loss}}} dV\, \varepsilon''(\mathbf{r})\,|\mathbf{E}|^2
=
-
\int_{V_{\mathrm{gain}}} dV\, \varepsilon''(\mathbf{r})\,|\mathbf{E}|^2,
\end{equation}
thereby enabling, in principle, the complete suppression of internal dissipation. In addition, the functional may even become  negative, signaling an effective gain capable of radiation loss compensation while preserving the bound state in continuum character of the mode.

\textbf{Finite-momentum non-Hermitian linewidth collapse in the $\mathcal{PT}$-symmetric regime.}
\noindent 
We considered a $\mathcal{PT}$-symmetric version of the photonic crystal slab examined above, case $\Delta/\Delta_{\mathrm{EP}}=0.1$ ($a=410$ nm, $h = 183$ nm, $r= 175$ nm) obtained by introducing balanced gain and loss within the unit cell. The complex permittivity satisfies the symmetry relation
\(\varepsilon(\mathbf r)=\varepsilon^\star(-\mathbf r)\),
implemented by assigning equal-magnitude gain and loss to opposite halves of the unit cell. In this configuration, the imaginary part of the dielectric function becomes sign-indefinite:
\(
\varepsilon''(\mathbf r)>0\) in loss regions, and
\(
\varepsilon''(\mathbf r)<0
\) in gain regions,
allowing the nonradiative contribution to the modal linewidth to undergo exact compensation.
The total modal quality factor can be expressed as
\begin{equation}
\frac{1}{Q_{\mathrm{tot}}}
=
\frac{1}{Q_{\mathrm{rad}}}
+
\frac{1}{Q_{\mathrm{abs}}},
\end{equation}
or equivalently in terms of decay rates,
\begin{equation}
\gamma_{\mathrm{tot}}
=
\gamma_{\mathrm{rad}}
+
\gamma_{\mathrm{abs}}^{\mathrm{eff}},
\end{equation}
where $\gamma_{\mathrm{abs}}^{\mathrm{eff}}$ represents the effective nonradiative contribution, including both dissipative and amplifying channels.
In passive quasi-BIC systems, the divergence of the quality factor follows the well-known symmetry-protected scaling law
\(
Q_{\mathrm{rad}}
\propto
k_{\parallel}^{-2},\)
which originates from the quadratic opening of the radiative channel under in-plane symmetry breaking as $k_{\parallel}\rightarrow 0$. In that regime, the divergence is governed by the progressive suppression of the coupling to the radiation continuum at the $\Gamma$ point.

\begin{figure}[h!]
    \centering
    \includegraphics[width=1\linewidth]{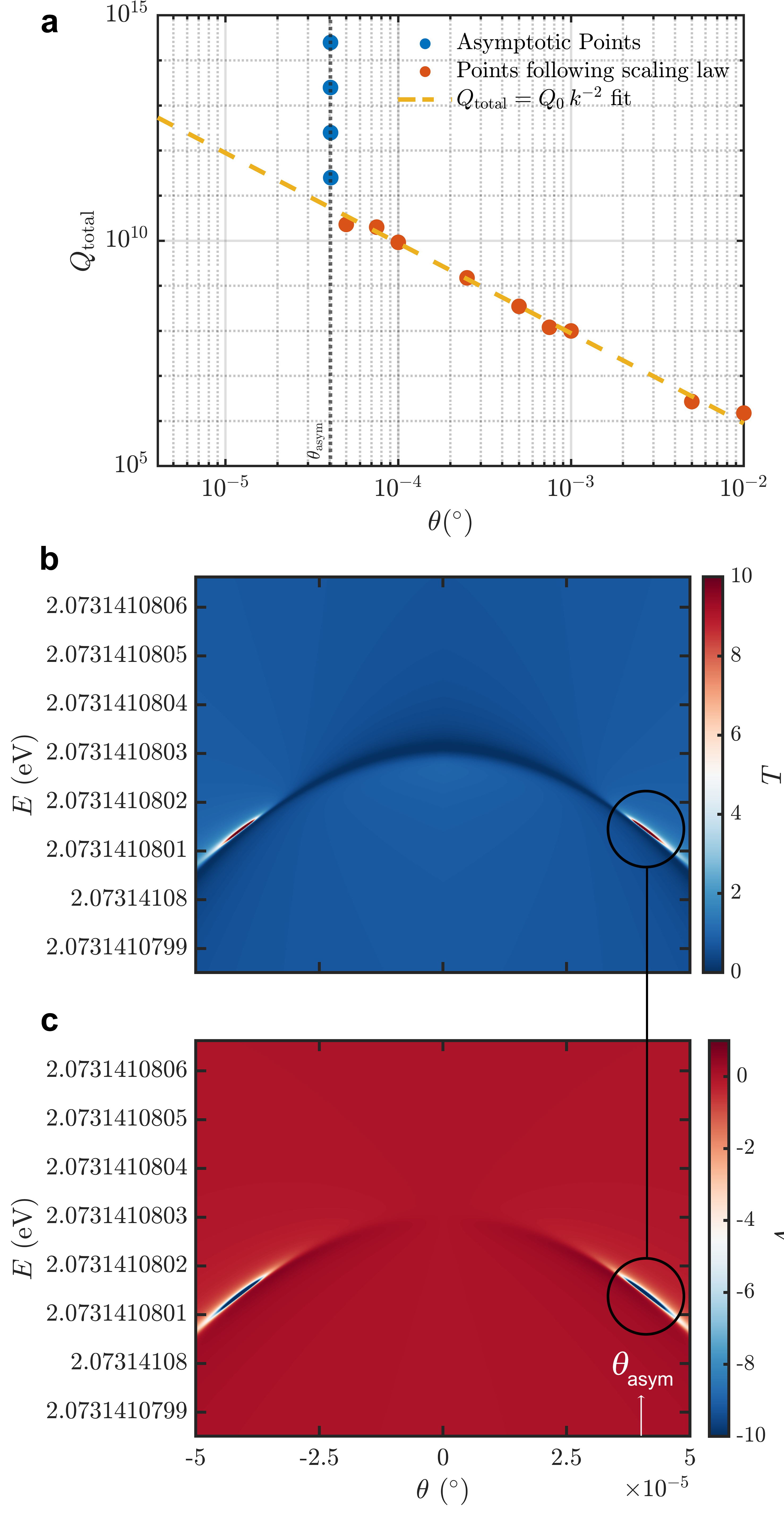}
    \caption{
(a) Total quality factor $Q_{\mathrm{tot}}$ extracted from RCWA simulations as a function of the incidence angle $\theta$ for the balanced gain--loss configuration. Away from the singularity, the resonance follows the conventional quasi-BIC scaling law $Q_{\mathrm{tot}}\propto k_{\parallel}^{-2}$ (orange points and dashed fit). Near the finite asymptotic momentum $\theta_{\mathrm{asym}}$, the scaling law breaks down and $Q_{\mathrm{tot}}$ exhibits a vertical asymptote, revealing a finite-momentum linewidth collapse.
(b) Angularly resolved transmittance spectrum $T(E,\theta)$ and
(c) corresponding absorption map $A(E,\theta)=1-T-R$, displaying a resonant amplification (negative absorption) at the same finite momentum where the quality factor diverges. Please note that $A\to - 10^{4}$ (i.e. effectively towards $-\infty$), but the colormap has been truncated for better visibility.}
    \label{fig:11}
\end{figure}

Remarkably, the present $\mathcal{PT}$-symmetric configuration exhibits a qualitatively different behavior, as shown \textbf{in Fig.~\ref{fig:11}a}. RCWA simulations reveal that the divergence of the total quality factor does not simply occur as $k_{\parallel}\rightarrow 0$. Instead, at a finite momentum
\begin{equation}
k_{\parallel}=k_{\mathrm{opt}}\neq 0,
\end{equation} $Q_{\mathrm{tot}}$ increases by several orders of magnitude while $k_{\parallel}$ remains nearly constant ($k_{\mathrm{opt}} = k_{0}\sin\theta_{\mathrm{asym}}$). The resulting behavior manifests as a vertical asymptote in the quality factor, indicating the collapse of the total modal linewidth at finite momentum,
\begin{equation}
\gamma_{\mathrm{tot}}(k_{\mathrm{opt}})
\rightarrow
0.
\end{equation}
This observation demonstrates that the divergence is not governed by the conventional algebraic radiative scaling law associated with symmetry-protected BICs. Instead, the system approaches a finite-momentum non-Hermitian compensation point satisfying
\begin{equation}
\gamma_{\mathrm{abs}}^{\mathrm{eff}}(k_{\mathrm{opt}})
=
-
\gamma_{\mathrm{rad}}(k_{\mathrm{opt}}),
\label{eq:linewidth_compensation}
\end{equation}
a new general regime where the total linewidth vanishes despite the persistence of finite radiative coupling. The absorption spectrum develops a symmetric Lorentzian-shaped negative peak.  In the amplifying regime, negative absorptance,
\(
A = 1 - T - R < 0,
\) indicates net extraction of energy from the gain medium into the electromagnetic field.\\
\indent As the system approaches the singular condition, the effective non-Hermitian linewidth
progressively collapses toward zero, \(\gamma_{\mathrm{tot}}\to0\),
and is thus accompanied by a divergence of the resonant total quality factor \(Q_{\mathrm{total}}\to \infty\), which demonstrates the existence of generalized BICs not only as a mathematical entity but directly through  thorough RCWA numerical simulations of PhCSs. 
The gBIC mode profile at $\theta_{\mathrm{asymptote}}$ projected on the zero order diffraction channel plane wave allows to extract $\gamma_{\mathrm{rad}}$. Calculating the mode profile loss integral on the $\mathcal{PT}$-symmetric domain, and extracting similarly $\gamma_{\mathbf{abs}}=\zeta_{2}$ at the gBIC point, reveals exactly the balance achievement within numerical accuracy: $\gamma_{\mathbf{rad}}=-\gamma_{\mathbf{abs}}^{\mathrm{eff}}=-\zeta_2$.
Thereby, the scattering response develops a pole-like amplification singularity, causing the absorptance to diverge toward large negative values \( 
A \rightarrow -\infty\).\\
\indent Unlike the previously discussed non-Hermitian mode redistribution in the passive system where absorption interference reshapes the mode profile, here the spatial field profile remains invariant across this transition, remaining apparently identical as \textbf{Fig.~\ref{fig:10Rev}a} panel (i), but with diverging maximum intensity, limited only by the finite computational resolution to $\sim10^{14}$. 
 The present system exhibits no substantial modal delocalization or absorption-channel reshaping. Instead, the compensation mechanism occurs directly at the level of the modal eigenvalue while leaving the eigenvector essentially unchanged.
The nonradiative term
converges to a finite negative value balancing effectively  the radiation loss, thereby allowing energy injection through the quasi-BIC radiative coupling at finite momentum.\\
\indent  Finally, the finite momentum value at which this singularity occurs depends on the overlap of the mode with the lossless and loss-gain regions of the resonator. Indeed it occurs at larger momentum when the field is well inside the loss-gain slab (because less evanescently delocalized for increasing $\Delta/\Delta_{\mathrm{EP}}$). Finally, we notice that the finite angle ($k_{\mathrm{opt}}$) of divergence tends to the supercritical coupling value $\theta_{\mathrm{sc}}^{\star}$ established when the material-limited absorption quality factor is $Q_{\mathrm{abs}}\simeq 10^{10}$, which will be further discussed elsewhere.  \\  \indent It is also worth mentioning that such a $\mathcal{PT}$-symmetric PhCS (or any analogous version) could be fabricated by doping the unit cell with a gain medium, such as lanthanide ions under external optical pumping.

\textbf{Summary.} For what concerns passive resonators, our results identify a regime in which BIC-based field enhancement is governed by the interplay of non-Hermitian modal structure and spatial field distribution. Both radiative and absorptive channels can be redistributed through modal interference and symmetry engineering. In gain--loss balanced systems instead, spectral compensation opens a completely new scenario that here we introduce with the concept of a generalized BIC, and that will deserve further investigation.\\
\indent More broadly, this perspective aligns with the emerging paradigm of coherent loss engineering in functional photonic platforms, where structured metasurfaces are designed to implement specific operators on optical fields~\cite{esfahani2025nonlocal}.

\section{Discussion}

\noindent
We have established a non-Hermitian framework in which quasi-BIC excitation is governed by the combined structure of radiation, reactive coupling, and absorption. In the bright--dark basis, the Friedrich--Wintgen condition appears as the singular limit of a rank-deficient radiation operator, while any finite leakage of the quasi-dark state induces a causal reactive coupling \(\kappa'\). This coupling enables energy transfer into the dark sector and yields the optimal supercritical condition \(\kappa_{\mathrm{sc}}=\sqrt{\Gamma_b\Gamma_d}\), extending critical coupling to weakly radiative, high-\(Q\) states.\\
\indent
For Dirac-like dispersions, this condition defines an open-Dirac singularity: the Dirac gap becomes comparable to the supercritical scale, the quasi-BIC is optimally excited at finite momentum, and the system approaches a ring of exceptional points. RCWA confirms that in this regime the conventional criterion \(Q_{\mathrm{rad}}=Q_{\mathrm{abs}}\) fails, while the supercritical model correctly predicts the enhancement angle and field buildup.\\
\indent
A striking consequence is that absorption ceases to behave as a fixed scalar material parameter. Near the supercritical Dirac regime, two mechanisms act together. First, non-orthogonal eigenvectors generate absorptive cross-coupling, opening a coherent absorption channel and redistributing dissipation between modes by reshaping their overlap with the lossy material regions. Second, and dominantly in the present passive structures, the quasi-BIC delocalizes toward low-loss regions as the light line is approached, reducing its overlap with the absorbing slab and producing the scaling \(Q_{\mathrm{abs}}^{\mathrm{eff}}\propto1/\sqrt{k_\parallel-k_{\mathrm{ll}}}\).
The quasi-BIC may thus remain absorptively brightest in absolute intensity while exhibiting an enhanced effective quality factor.\\
\indent Intrinsic absorption becomes a dynamically engineered quantity controlled by modal structure and spatial redistribution. This motivates the introduction of the gBIC concept as a limiting state where radiative and absorptive decay are jointly minimized.  Engineered gain--loss landscapes enables exact cancellation of effective net gain and weak radiative coupling feeding the mode sustained by that gain. \\
\indent The observed finite-momentum singularity can be interpreted as a non-Hermitian critical balance between radiative leakage and momentum-induced effective gain. 
In conventional critical coupling $\gamma_{\mathbf{rad}}=\gamma_{\mathbf{abs}}$,
leading to maximal absorption,
vanishing reflection/transmission channel,
and finite field enhancement.
In this case instead, we have $\gamma_{\mathbf{rad}}=-\gamma_{\mathbf{abs}}$, which implies linewidth collapse, diverging lifetime, diverging intracavity field.  At exact $\Gamma$,
the field distribution samples gain and loss symmetrically,
so the net gain compensation may remain incomplete,
or constrained by symmetry. At finite momentum (quasi-BIC in the radiative only sense) the Bloch phase introduces a slight asymmetry and
the overlap with gain/loss regions becomes imbalanced,
producing an effective negative linewidth contribution. Even if the spatial profile appears visually similar, a tiny phase-weighted imbalance
is sufficient to shift the eigenvalue imaginary part dramatically when the total linewidth is already very small. So the singularity may emerge from weak momentum-induced overlap asymmetry,
amplified by the quasi-BIC high-Q background. In realistic systems, additional channels including gain saturation, disorder-induced scattering, nonlinear absorption, and spontaneous emission are expected to regularize the divergence, replacing the mathematical singularity with a finite yet potentially very large enhancement.\\
\indent
These results define a broader paradigm of non-Hermitian loss engineering, in which a photonic resonator is conceived as an open system whose dissipation is deterministically reshaped by its electromagnetic modal coupling. Topology, interference, and spatial delocalization jointly structure radiative and absorptive channels, elevating loss from a material constraint to a controllable design parameter that ultimately sets the performance limits of open photonic resonators.

\vspace{20pt}
\textbf{Acknowledgements.} We warmly thank Domenico Passaro (Institute of Applied Sciences and Intelligent Systems, National Research Council, Italy) for assistance with simulation infrastructure; and gratefully acknowledge Dr. Kirill Koshelev (Australian National University, Canberra) for insightful input. B.M. acknowledges the EU Italian National Recovery and Resilience Plan (NRRP) of NextGenerationEU (PE0000023-NQSTI). S.B.  acknowledges the EU Italian National Recovery and Resilience Plan (NRRP) of NextGenerationEU (PE00000001-RESTART).  S.R. acknowledges financial support by B-PLAS grant no. PRIN-2022TXWE32, MIUR, Italy.  G.Z. acknowledges financial support from the European Union – NextGenerationEU, under the National Recovery and Resilience Plan (NRRP), Mission 4, Component C2 Investment 1.1, PRIN project INSPIRE, grant no. P2022LETN5. 

\vspace{12pt}
\textbf{Author contributions:} GZ conceived the theoretical model. SB and BM carried out numerical simulations and analysed the data, with contribution from  SR and GZ. All authors discussed the results. GZ wrote the draft with contributions from SB, BM, and SR. SR and GZ edited the final manuscript. SR and GZ supervised the research.

\vspace{12pt}
\textbf{Competing interests:} The authors declare no competing interests. 

\vspace{12pt}
\textbf{Data availability.} All relevant data that support the findings of this work are available from the authors.

\vspace{12pt}
\textbf{Correspondence and requests for materials} should be addressed to Gianluigi Zito.

\bibliography{library}

\end{document}